\documentclass[a4paper]{biophys}
\usepackage{helvet,times,mathptmx}
\usepackage{bm,textcomp}

\gridframe{N}
\cropmark{N}

\usepackage{amsmath,amssymb,bm,graphicx,color}
\usepackage{subfig}

\usepackage[round,numbers,sort&compress]{natbib}

\usepackage{hyperref}
\hypersetup{
  pdfinfo={
    Title={Interplay of packing and flip-flop in local bilayer deformation. How phosphatidylglycerol could rescue mitochondrial function in a cardiolipin-deficient yeast mutant.},
    Author={Nada Khalifat, Mohammad  Rahimi, Anne-Florence  Bitbol, Michel Seigneuret, Jean-Baptiste Fournier, Nicolas Puff, Marino Arroyo, Miglena I. Angelova},
    Subject={},
    Keywords={}
  }
}
\usepackage{pdfpages}

\begin{document}
\renewcommand{\floatpagefraction}{.9999} 
\renewcommand{\textfraction}{.0001}
\setcounter{totalnumber}{8}
\renewcommand{\topfraction}{.9999}
\renewcommand{\bottomfraction}{.9999}

\renewcommand{\refname}{Supporting References} 
\renewcommand{\thefigure}{S\arabic{figure}}

\setcounter{page}{1} 



# Interplay of packing and flip-flop in local bilayer deformation. How phosphatidylglycerol could rescue mitochondrial function in a cardiolipin-deficient yeast mutant.

Nada Khalifat[†‡], Mohammad Rahimi[§], Anne-Florence Bitbol[§∥], Michel Seigneuret[**] [*], Jean-Baptiste Fournier[**], Nicolas Puff[**∥], Marino Arroyo[††], Miglena I. Angelova[**∥] [*]

[†]UPMC Univ Paris 06, UMR 168, Institut Curie, 26 rue d'Ulm, 75248 Paris Cedex 05, France
[‡] CNRS, UMR 168, Institut Curie, 26 rue d'Ulm, 75248 Paris Cedex 05, France
[§]Lewis-Sigler Institute for Integrative Genomics, Princeton University, Princeton, NJ 08544, USA
[∥]Department of Physics, Princeton University, Princeton, NJ 08544, USA
[**]Université Paris Diderot-Paris 7, Matière et Systèmes Complexes CNRS UMR 7057, Paris, France
[∥]Université Pierre et Marie Curie, Paris, France
[††]Departament de Matemàtica Aplicada III, LaCàN, Universitat Politècnica de Catalunya, 08034, Barcelona, Spain.

[*]Correspondence : miglena.anguelova@upmc.fr or michel.seigneuret@univ-paris-diderot.fr
Université Paris Diderot-Paris 7
Matière et Systèmes Complexes
CNRS UMR 7057
10 rue Alice Domon et Léonie Duquet
75205 Paris cedex 13
France

## ABSTRACT

In a previous work, we have shown that a spatially localized transmembrane pH gradient, produced by acid micro-injection near the external side of cardiolipin-containing giant unilamellar vesicles, leads to the formation of tubules that retract after the dissipation of this gradient. These tubules have morphologies similar to mitochondrial cristae. The tubulation effect is due to direct phospholipid packing modification in the outer leaflet that is promoted by protonation of cardiolipin headgroups. Here we compare the case of cardiolipin-containing giant unilamellar vesicles with that of phosphatidylglycerol-containing giant unilamellar vesicles. Local acidification also promotes formation of tubules in the latter. However, compared to cardiolipin-containing giant unilamellar vesicles the tubules are longer, exhibit a visible pearling and have a much longer lifetime after acid micro-injection is stopped. We attribute these differences to an additional mechanism that increases monolayer surface imbalance, namely inward PG flip-flop promoted by the local transmembrane pH-gradient. Simulations using a fully non-linear membrane model as well as geometrical calculations are in agreement with this hypothesis. Interestingly, among yeast mutants deficient in cardiolipin biosynthesis, only the *crd1*-null mutant, which accumulates phosphatidylglycerol, displays significant mitochondrial activity. Our work provides a possible explanation of such a property and further emphasizes the salient role of specific lipids in mitochondrial function.

## INTRODUCTION

The shapes of cell membranes as well as their global or local modifications have an essential role in cell function. The shape of a membrane, i.e. its local or global curvature, is related to the asymmetry of its composition (transbilayer asymmetry of lipids and proteins) and its environment (ion gradients, unilateral binding of effector molecules). Many studies have attempted to mimic such shape changes in lipid bilayers (1-3). Membrane shape/curvature changes often occur through local or global modifications of the relative area of the two bilayer leaflets (3-5). Such processes may be associated with perturbations of the lipids per se. In this regard, two main mechanisms have been documented: (i) direct differential change in the packing (i.e. the effective molecular area and/or shape) of specific lipid species between the two monolayers due to e.g. asymmetrical ligand or protein binding, chemical modification or electrostatic effects (6-12); (ii) asymmetric lipid transbilayer diffusion, either spontaneous or protein mediated (13-15). In the present study, we describe an example of a local shape change of a lipid bilayer where there is interplay of these two processes, in the context of a model membrane system that mimics the mitochondrial inner membrane of a yeast cardiolipin (CL)-deficient mutant which accumulates phosphatidylglycerol (PG).

The mitochondrial inner membrane (MIM) is a salient case in which bilayer shape and its modifications are strongly coupled with organelle function. It is the main site for ATP production in aerobic eukaryotes, associated with a transmembrane electrochemical proton gradient. This gradient couples an $O_2$-fuelled respiratory chain with an ATP-synthase, as explained by the chemiosmotic theory. Such processes take place mainly in tubular invaginations of the MIM, termed cristae (16). Their particular local shape appears to be strongly coupled with function since respiratory rate and ATP synthesis affect cristae morphology (17). Depending on the physiological state of the mitochondria, the mitochondrial matrix shape can be either "condensed", with thick cristae, or "orthodox", with thinner cristae. The origin of the MIM cristae shape as well as of its coupling to mitochondrial function is the subject of intense research. CL, a particular lipid found almost only in the MIM, likely plays an important role since abnormalities in its synthesis usually lead to drastic changes in organization of the MIM, loss of ATP synthesis as well as of the cristae morphology (18). However, several authors (19-22) have reported that the yeast *crd1*-null mutant, which lacks CL synthase and has no detectable CL but accumulates phosphatidylglycerol (PG), displays normal mitochondrial activity at 30°C in glucose and non-fermentable carbon sources, although mitochondrial functional defects are apparent under suboptimal substrate conditions and higher temperature. This suggests that PG can specifically substitute for some functions of CL in the MIM.

In a previous work, using giant unilamellar vesicles (GUVs) made of phosphatidylcholine (PC)/phosphatidylethanolamine/CL or PC/CL to model the MIM, we offered some new insights into the factors that determine the dynamical tubular structures of the mitochondrial inner membrane cristae (23, 24). We found that when such vesicles are locally submitted to an external local acidification, by topically generating a higher pH outside of the membrane, inward tubulations develop. The morphology of these tubules is strongly reminiscent of MIM cristae. This suggests that cristae formation can be accounted for within a minimal purely lipidic model membrane under the influence of an electrochemical proton gradient. Additionally, we have developed a theoretical description of local membrane deformations due to local pH variations (11, 12, 23-26). The asymmetric pH-induced changes of protonation of charged headgroups and the resulting changes of the electrostatic interactions between such headgroups promote asymmetric local modifications of the lipid packing. This affects both equilibrium lipid density and monolayer local spontaneous curvature, resulting in

a local deformation of the membrane (12, 26). Intermonolayer friction plays an important role in the relaxation dynamics of these deformations.

In the present study, we further refine this model by focusing on yeast mutants which feature CL-deficient but PG-enriched membranes. Using PC/PG-GUVs, we find that a local pH-induced tubulation also occurs but with a different morphology and dynamics. We show that such behavior can be accounted for in the framework of our model, provided that specific properties of PG regarding flip-flop are taken into account. These results provide an explanation of the functionality of the *crd1*-null yeast mutant.

**MATERIALS AND METHODS**

**Chemicals**

Lipids were purchased and used without further purification. Egg yolk L-α-phosphatidylcholine (PC) from Sigma-Aldrich, heart bovine 1,3-bis(sn-3-phosphatidyl)-sn-glycerol diphosphatidylglycerol (CL) and L-α-phosphatidylglycerol (PG) from Avanti Polar lipids. Fluorescent probes 6-dodecanoyl-2-dimethylaminonaphthalene (Laurdan) from Molecular Probes, HEPES from Interchim and EDTA from Sigma-Aldrich.

**Giant Unilamellar Vesicles Preparation**

GUVs were made on platinum electrodes in 2ml of buffer A (pH 8, HEPES 0.5 mM, EDTA 0.5mM) in a temperature controlled chamber (T ~ 25 °C) by the electroformation method as described in (23). The GUVs were used the same day. Electroformation is known to produce vesicles that are ~90% unilamellar (27, 28). Unilamellarity of studied vesicles was checked by careful phase-contrast microscopy examination of the junction with the electrode as well as of long-wavelength thermal membrane fluctuations.

**Video Microscopy and Microinjection**

The GUVs were observed by phase contrast with a Zeiss Axiovert 200M microscope, equipped with a charge-coupled device camera (CoolSNAP HQ, Photometrics, Tucson, AZ, USA). The experiments were computer-controlled using Metamorph software (Molecular Devices, Downington, PA, USA).

Tapered micropipettes for the local injection of HCl were made from GDC-1 borosilicate capillaries (Narishige, Tokyo, Japan), pulled on a PC-10 pipette puller (Narishige). For local injection of the acidic solution of HCl (100 mM, pH 1.6) a microinjection system (Eppendorf femtojet) was used with capillary of 0.3 μm and injection pressure set at 15–25 hPa. The positioning of the micropipettes was controlled by a high-graduation micromanipulator (MWO-202, Narishige) and the initial distance from the GUV membrane was ~10 μm.

**Large Unilamellar Vesicle Preparation**

Large unilamellar vesicles (LUVs) were prepared in buffer B (HEPES 5 mM, EDTA 0.1 mM, pH 8), using the extrusion method described in (24). The fluorescent probe Laurdan was mixed with the lipids in the initial organic solution and was used at low concentrations in order not to perturb the lipid bilayer structure at 1/200 Laurdan/lipid ratio. The LUVs solution were kept at 4 °C, and used for measurements the day after.

3

**Laurdan Fluorescence Measurements**

Steady-state fluorescence measurements were carried out with a Cary Eclipse spectrofluorimeter (Varian Instruments, CA) equipped with polarizers and a thermostated cuvette holder (±0.1 °C). Excitation and emission slits were adjusted to 5 nm. Fluorescence emission spectra were all recorded at 25 °C. All fluorescence measurements were carried out at a total lipid concentration of 0.2 mM.
The pH of the LUV samples was gradually lowered by adding aliquots of 10 or 100 mM HCl. Measurements were made after 5 min of equilibration under agitation. Excitation wavelength was 355 nm. Shifts in the Laurdan emission spectrum were quantified by calculating the generalized polarization (GP) defined as GP = $(I_{440} − I_{490})/(I_{440} + I_{490})$, (29)

**RESULTS AND DISCUSSION**

**Comparison of tubular shape invaginations under local pH modulation in CL- and PG-containing GUVs.**

Our previous studies have shown that a localized change in pH in the vicinity of a GUV can elicit a local shape transformation of the bilayer, depending on the lipid composition (11, 12, 23-25). Such a pH sensitivity, requires the presence of a lipid species, which undergoes a change of its degree of protonation and therefore of its overall charge. The mechanism we have proposed and detailed theoretically is as follows. The local variation of the average charge leads to a local modification of the mutual repulsion between lipid molecules in the outer monolayer. This directly affects their packing, by changing both their effective molecular shape and their equilibrium in-plane area. This promotes a tendency for local shape change since the outer monolayer of the exposed membrane zone has a new spontaneous curvature and a new equilibrium molecular density. There is therefore a local mechanical asymmetry between the two monolayers. A relaxation mechanism is the relative sliding of the two monolayers that tends to spread the local density mismatch (although not the spontaneous curvature mismatch) between these monolayers over the whole membrane. However, since this process is slow, due to intermonolayer friction, a transient shape change occurs to accommodate the asymmetry between the two monolayers. The exact nature and magnitude of the shape change depends upon the sign and the strength of the initial mechanical stimulus caused by the chemical modification and upon other factors (such as the GUV membrane tension). This is the mechanism that we have previously invoked for the formation of cristae-like tubulations in CL-containing GUVs.
In the present study, we have compared the effect of a local pH gradient on GUVs with lipid composition corresponding to that (i) of normal IM membrane and (ii) of the yeast *cdr1*-null mutant devoid of CL and accumulating PG. For this purpose, we have used a setup similar to that of our previous study. Namely, individual GUVs in buffer at pH 8 (similar to the mitochondrial matrix pH) were locally submitted to an external pH decrease by continuous acid microinjection in the vicinity of the membrane (distance 10 μm) to mimic locally acidification due to the activity of the mitochondrial respiratory chain. In our previous work (23), we roughly estimated that such acidification brings the bulk pH near the GUV to a value of 4-5 by a calculation based on visualizing the acid flux from the micropipette under the microscope. Here, estimation of the steady-state pH at the membrane surface was made using a more precise theoretical description of the proton flux, taking into account the actual micropipette geometry and realistic values for the injection pressure range of our experiments (see Supporting Material). This calculation yielded a pH value of 4.8±0.4 at the membrane

surface facing the micropipette for a 50 μm diameter vesicle and for a 30 s injection. This is consistent with our earlier estimation. Two lipid compositions PC/CL 90:10 and PC/PG 90:10 (mol/mol) were compared. Below are shown representative data for the two cases.

As already reported previously, in the case of CL-containing GUVs, such local pH decrease promotes rapid and progressive formation of inward tubular invaginations (Supporting Material: Fig. S2, movie S1). The morphology of these tubes strongly recalls mitochondrial cristae (23, 24). When acid microinjection is stopped, the tube retracts (*frames from t= 31.8 s to 71.7 s in* Fig. S2). Its retraction time depends upon the size and the membrane tension of the GUV and remains similar for comparable vesicles (see Fig 2 *C*). For the relatively large (diameter ~50 μm) and slightly deflated vesicle of Fig. S2, the retraction time is equal to about 40 s.

When PG is substituted for CL, local acidification also promotes tubulation with similar growth kinetics (Fig. 1, movie S2). However, two major differences are apparent. First, the morphology of these tubules is markedly different. On average, these grow longer than with CL and ultimately exhibit a pearled morphology. Second, after acid injection is switched off (*frames from t=27.3 s to t=203.3 s* on Fig. 1), the kinetics of tube retraction is significantly slower.

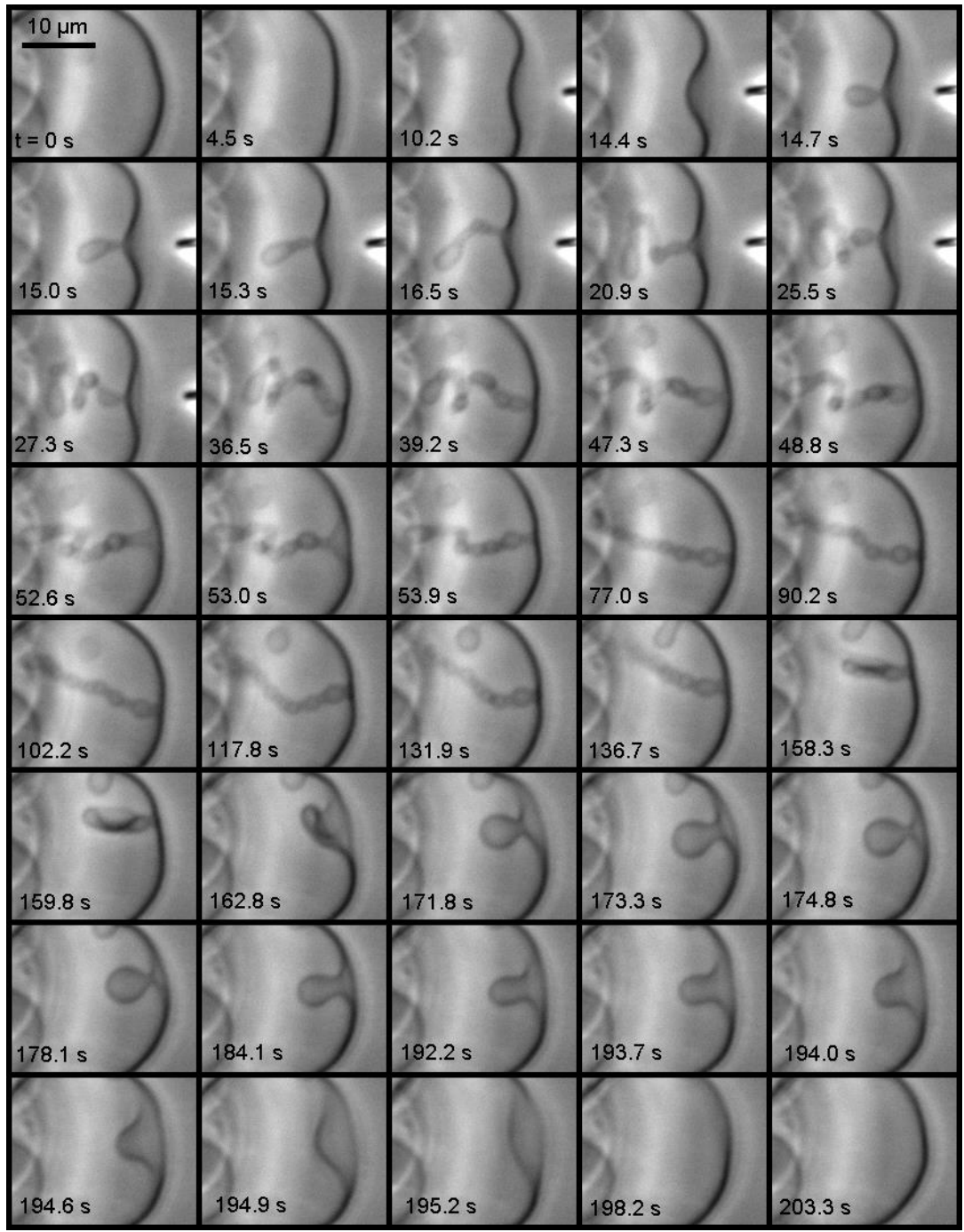

FIGURE 1 Local membrane invagination triggered on a GUV composed of PC/PG 90:10 mol/mol by a local pH decrease performed using a micropipette delivering an acidic solution (100 mM HCl, pH 1.6) in buffer A at pH 8. The acid microinjection is started at $t=0$ and ended at $t=27.3\ s$. The corresponding movie is movie S2.

In the case of the vesicle of Fig. 1, which has a similar size and tension than that of Fig. S2, the time for tube retraction is 170 s, i.e. more than 4 times slower that for CL-vesicles.

6

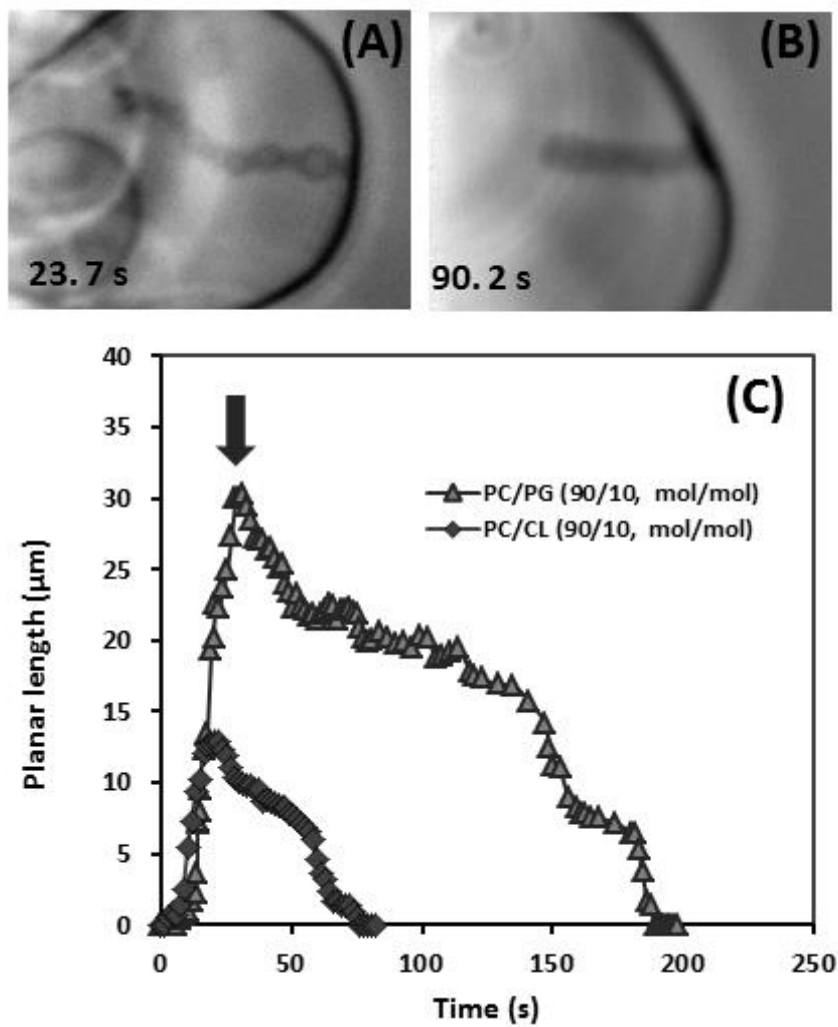

FIGURE 2 Comparison of the morphologies of tubes formed in GUVs of composition PC/PG 9:1 mol/mol (A) and PC/CL 9:1 mol/mol (B) upon local microinjection with acid. The images correspond respectively to *t= 23.7* and *t=90.2s* after injection. (C) Comparison of the kinetics of tube growth upon local microinjection with acid for GUVs of composition PC/CL 9:1 mol/mol (*diamonds*) and PC/PG 9:1 mol/mol (*triangles*). The apparent planar length of tubes was measured manually on microscopy images using the curvilinear tool of ImageJ. The microinjection starts at t=0 and ends where indicated by a vertical arrow.


Fig. 2 *A* and *B* compare the CL vesicle and PG vesicle tubules of Fig. S2 and 1, emphasizing their different morphologies. The different dynamical behavior is quantified in Fig 2 *C*, which shows the evolution of the planar length of the tubule (i.e. the length as measured on microscopic images which corresponds to a planar projection of the tubule).

The above data is representative of the behavior of large, slightly deflated, vesicles containing CL or PG. Local acidification of smaller and/or tenser vesicles formed extremely long, narrow and entangled tubules that retracted very slowly. A possible contributing effect on shape changes of the osmotic pressure difference generated by acid injection was ruled out by submitting GUVs to injection of 100 mM or higher sucrose under similar conditions: no shape change was observed (not shown).

**pH effect on lipid packing in PG- and CL-containing LUVs**

To get more insight into the mechanism of the local pH-induced tubulation in PG containing bilayers, as compared to CL-containing bilayer, we have studied the effect of pH on lipid packing using the fluorescence of Laurdan incorporated in the membranes of LUVs of similar lipid composition as the GUVs described above. Besides providing useful information about the effect of pH and headgroup protonation state on lipid packing, this data allows one to estimate of the "apparent $pK_a$" of such protonation phenomena. In bilayers containing charged lipids, and due to the negative surface potential, the $H^+$ concentration in the vicinity of the membrane is likely to be different than in the bulk. Charged lipids, depending on their concentration in the membrane modulate the pH range for their own protonation. The extent of this modulation also depends on ionic strength (30).

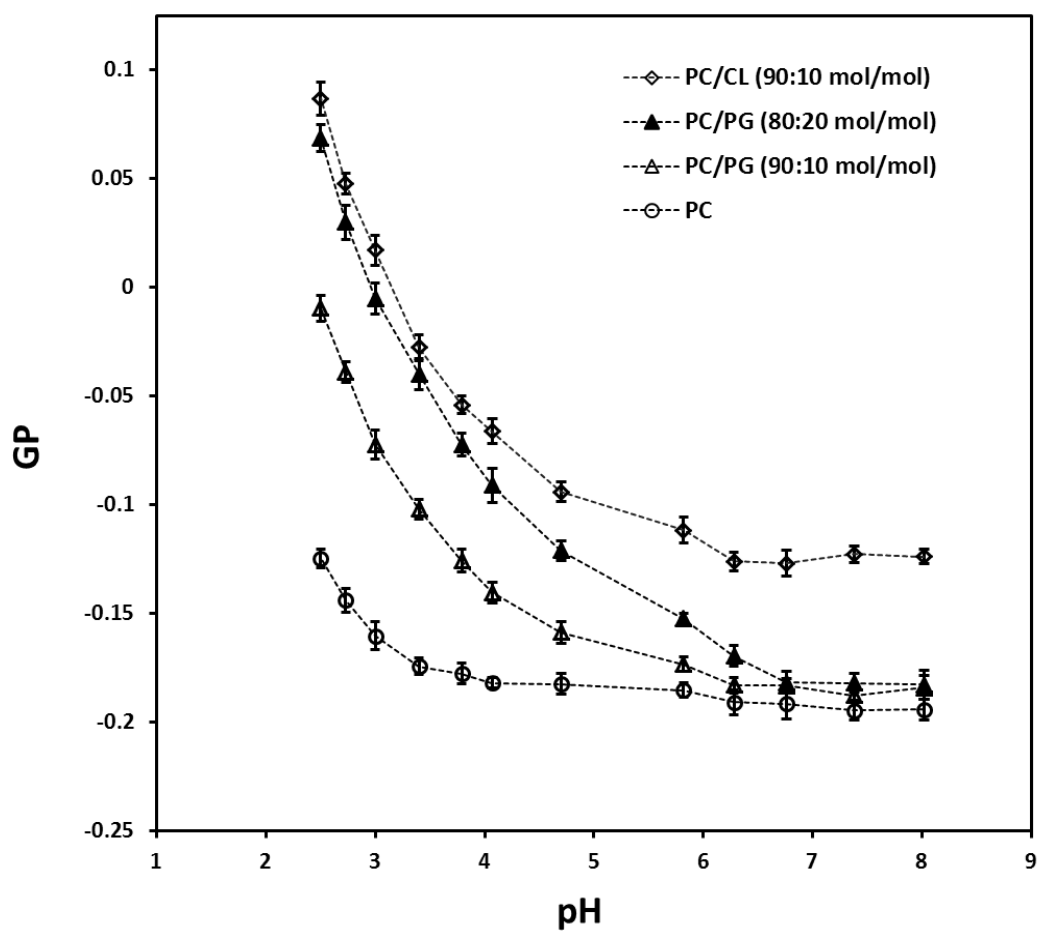

FIGURE 3 Effect of pH on the general polarization measured from fluorescence emission spectra of Laurdan in LUVs of various compositions at 25°C. The Laurdan /lipid ratio is 1:200.(mol/mol) and the GP is calculated as $GP = (I_{440} - I_{490})/(I_{440} + I_{490})$.

Fig. 3 shows the effect of pH on the Laurdan generalized polarization for various LUVs composition. For PC/CL 90:10, decreasing pH gives rise to an important increase of GP and therefore of lipid packing in the hydrated region of the membrane. This is in agreement with our previous studies (1-3). This results from the protonation of CL headgroup which

8

decreases the electrostatic repulsion between CL molecules and therefore changes their effective molecular shape as well as their molecular surface area. We previously showed that such effects are the effective driving force for the pH-induced shape transformations found in CL-containing GUVs (24). The pH effect on CL bilayers occurs in the bulk pH range 6-2.5 suggesting that this is the actual range for the CL headgroup protonation under such conditions. When PG is substituted for CL in identical proportion (PC/PG 90:10), a rather similar effect is observed on GP when pH is lowered, although the overall GP values are higher with CL, due to the well-known ordering effect of the CL backbone (31). The pH range for the effect is also similar. No such pH-induced packing variation occurs in this range for the zwitterionic phosphatidylcholine, emphasizing its specificity for charged phospholipids.

For 10 % PG, the increase in packing occurs between pH 7 and 2.5. This is to be compared with a reported intrinsic pK value of 2.9 for PG (32). However, when the proportion of PG in GUVs is doubled to 20%, the low-pH induced increase in GP is shifted by 0.5-1 pH units towards higher pH. This illustrates the mutual influence of PG molecules on their protonation behavior, namely, the higher the PG percentage in the membrane, the more its protonation is anti-cooperatively hindered due to the unfavorable mutual electrostatic repulsion as well as to the increased membrane charge that attracts H+ towards the surface (30). Thus the deprotonation of negative lipids in membranes can be strongly shifted or extended to higher pH.

These data indicate that, for PG-containing GUVs, a shape-changing mechanism similar to that proposed for CL-containing GUVs (23, 24) is operative. Indeed, at 10% PG, a pH decrease from 8 to 4.8±0.4, identical to that imposed in the GUV experiments, promotes a significant increase in packing due to increase in molecular equilibrium density and monolayer spontaneous curvature. It is therefore likely that such a mechanism contributes to the tubulation observed for PG-containing GUVs.

**Geometrical interpretation of the pearled tube morphology**

In spite of this similarity, significant differences are observed with regards to the local shape change elicited by local acidification between CL and PG containing GUVs. A particularly salient one is the different morphology of the tubes formed, which, in the case of PG, are longer and exhibit pearling. As shown from geometrical arguments in our previous work (23), a tube is a more efficient way of accommodating surface curvature as compared to, e.g. a bud. This suggests that the longer tube length obtained with PG is due to a higher surface curvature and therefore a higher lipid density mismatch between the two monolayers. In this section, we show that further geometrical arguments may explain why a lipid density mismatch may also lead to an undulated tube. We consider a membrane tube of radius $R_0$ and study whether the area-difference between its two monolayers increases or decreases when a peristaltic undulation is established. We choose a sinusoidal undulation that conserves the volume parameterized in cylindrical coordinates by

$$r(z) = R_o \left(1 + \frac{\varepsilon^2}{2}\right)^{-\frac{1}{2}} \left[1 + \varepsilon \cos\left(\frac{2\pi z}{\lambda}\right)\right] \quad (1)$$

where $\varepsilon$ is the amplitude of the undulation and $\lambda$ its wavelength. We do not assume $\varepsilon$ to be small. The area-difference between the two monolayers is proportional to the mean-curvature H of the membrane (4). The latter is given by

9

$$H(z) = \frac{1 + r'(z)^2 - r(z)r''(z)}{r(z)\left[1 + r'(z)^2\right]^{3/2}} \quad (2)$$

Where *r'* and *r''* are respectively the first and second z-derivatives of *r*. For given amplitudes and wavelength of the undulation, we have numerically calculated the average <*H*> of the mean-curvature (and hence the area difference) along the tube axis. As shown in Fig. 4, roughly around $\lambda \boldsymbol{.} 3R_0$, the average mean-curvature decreases as the tube undulates, whereas roughly around $\lambda/3R_0$ it increases, the increase being significant if the amplitude is larger than about 30%.

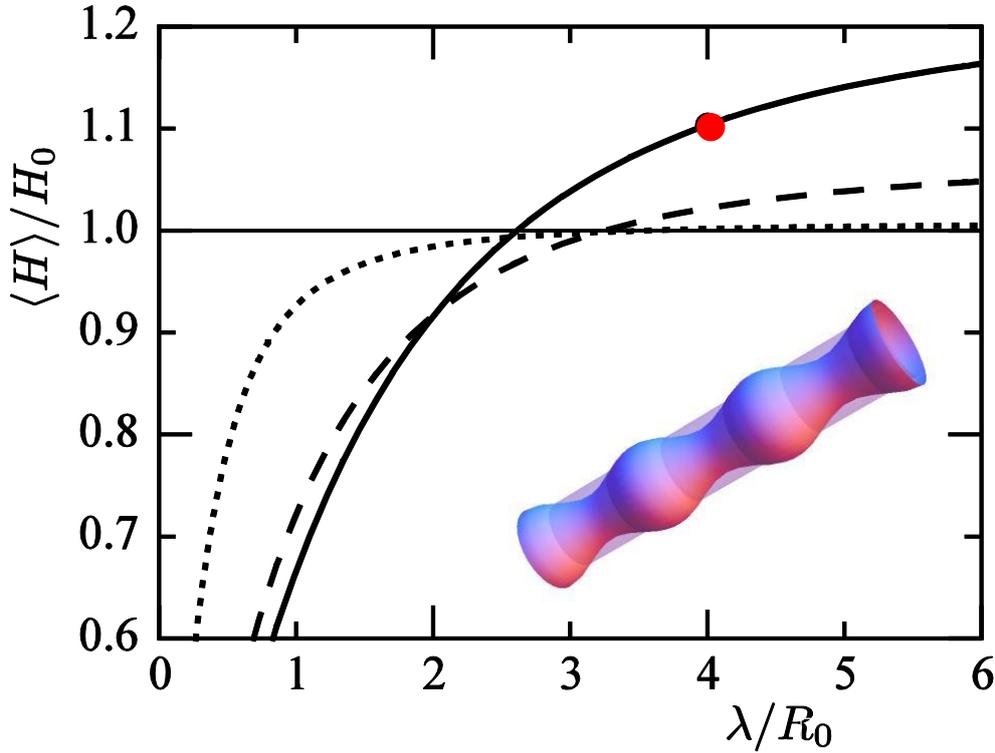

FIGURE 4: Average mean-curvature <H> of a tube subject to a peristaltic undulation (inset) normalized by the mean-curvature $H_0 = 1/2R_0$ of the tube of radius $R_0$ with the same volume, as a function of the wavelength $\lambda$ of the undulation. The area-difference between the two monolayers is proportional to the mean-curvature H of the membrane. The relative amplitude $\varepsilon$ of the undulation is equal to 0.1 (dotted line), 0.3 (dashed line) and 0.5 (plain line). The inset corresponds to the situation indicated by a red dot.

This shows that imposition of an area difference above a certain threshold can indeed induce pearling of a tube, as observed experimentally for PG-containing GUVs. Thus both the longer length and the pearling of the tubes formed in PC/PG-GUVs are due to the occurrence of a higher pH-induced monolayer mechanical imbalance as compared to CL-containing GUVs.

**Hypothesis on the differences of the tubulation processes of PC/CL- and PC/PG-GUVs.**

The above data suggest that, although the pH-induced tubulation in CL and PG vesicles share one common shape change mechanism, a second mechanism is also operative for PC/PG-GUVs which adds to the surface imbalance, hence the longer tube length and the pearling. This later mechanism is likely also responsible for the slower tube retraction. Interestingly, Cullis and collaborators (33, 34) demonstrated that PG can undergo rapid transbilayer diffusion in LUVs through its neutral form even above its pK. In the presence of a pH gradient, the formation of the neutral intermediate is more favorable in the monolayer facing the most acidic solution, so that more PG molecules leave this monolayer towards the other one than the opposite. Furthermore, after monolayer translocation, neutral PG molecules are bound to become quickly deprotonated in the new basic environment, making the reverse motion very unlikely. In short, protonated PG becomes neutral and can easily flip towards the higher pH leaflet driven by mixing entropy differences, but then by losing its proton in the new environment the reverse translocation is kinetically blocked. This results in a net transport of PG towards the monolayer and an asymmetric distribution of PG, with PG accumulating in the monolayer facing the most basic solution. No similar transbilayer diffusion is found for CL (33). This difference can be explained by the smaller size of the PG headgroup compared to the bulky CL headgroup and by that fact that PG can be brought to neutral form by binding of a single $H^+$ whereas CL requires binding of 2 $H^+$, a much rarer event.

Considering the well-known relation between phospholipid asymmetric flip-flop and membrane curvature (13-15), we propose that inward PG flip-flop occurs in the area exposed to acid in GUV experiments. Since the external surface of this area is more acidic, down to a pH close to 4, this generates a net inward transport of PG from the outer to the inner monolayer thereby creating an asymmetric distribution of PG. Such a process leads to a decrease of the outer monolayer density and an increase in the inner monolayer density and therefore to an additional density mismatch between the two monolayers and therefore to a further bilayer imbalance. This adds up to the direct mechanism involving electrostatic repulsion and modification of equilibrium density mismatch/spontaneous curvature. Interestingly, global shape changes of PG-containing GUVs upon bulk external pH acidification have already been reported (14, 35).

An a priori estimate of the relative efficiency of both mechanisms can be obtained as follows. According to (34), the half-time for PG flip-flop induced by acidification to pH 4 is 12 min at T=25°C. It can be straightforwardly calculated that, at 10% total PG, this corresponds to a PG translocation corresponding to a fraction $A \sim t.\ln2/t_{1/2}= 0.01$ of total lipid after 10 s in the acid-exposed area (10% of the total area) and therefore of 0.001 relative to total outer monolayer lipid. This corresponds to a relative area change $\Delta a/a$ between the two monolayers of 0.2%. This is similar to what we estimated for the condensation/spontaneous curvature mechanism in our earlier work (23). Therefore in the case of PG, two mechanisms of similar magnitude may contribute in synergy to the bilayer imbalance. It should be remarked that the occurrence of the PG flip-flop mechanism does not decrease the efficiency of the PG deprotonation/packing mechanism since the depletion of exterior PG due to flip flop is negligible. Therefore a longer length and pearling of tubules may be expected with PC/PG-GUVs compared to PC/CL-GUVs where only one mechanism is operative.

The occurrence of an acid-driven inward PG flip-flop for PC/PG-GUVs might also explain the slower rate of tube retraction after injection is stopped. Indeed, the interruption of acid injection is followed by dissipation of the local low pH inside the tube, limited by $H^+$

diffusion out of the tube. For PC/CL-GUVs, where protonation-related change in packing is the sole mechanism, such pH homogenization immediately promotes deprotonation of CL and relaxation of the membrane deformation (i.e. tube retraction), in order to recover the initial lipid density and monolayer curvature match between the two monolayers. Relative sliding of the two monolayers also contributes to relaxation but only minimally since it is slower than diffusion of the $H^+$ ions (12).

With PG, when injection is stopped, the above-mentioned relaxation mechanism associated with dissipation of the pH gradient and PG headgroup deprotonation is operative but can only yield partial tube retraction. Indeed, even when this gradient has collapsed, outward flip-flop of excess PG, although thermodynamically favorable is kinetically frozen. The rate of unidirectional flip-flop being proportional to the concentration of neutral PG in the leaflet and therefore to the $H^+$ concentration (34), the rate of outward flip-flop from the internal leaflet (pH 8) is much slower (in principle 1000-10000 times) than the rate of the previous inward flip-flop from the external leaflet (pH ~4-5), so that inside PG is effectively trapped in the internal leaflet. Therefore, the relaxation of the density asymmetry caused by asymmetric flip-flop only occurs through slow relaxation mechanisms. Monolayer relative sliding which delocalizes the residual density mismatch over the whole bilayer is a slow process. An estimation of the timescale can be obtained in the limit of small deformation as $\gamma_s = K q^2/2b_m$, $K_s = 0.1/\text{J}\cdot\text{m}^{-2}$ being the monolayer stretching modulus, $b_m = 10^9 \text{J}\cdot\text{s}\cdot\text{m}^{-1}$ the inter-monolayer friction coefficient and $q = \pi/2L$, the wavenumber of the deformation, with $L=30$ μm, the length of the tube (25). This corresponds to a timescale of about 30 s.

Lipid lateral diffusion is also able to relax the transbilayer density mismatches related to flip-flop and may contribute with a similar timescale i.e. $L^2/D_{lip} = 20$ s with a value of the lipid lateral diffusion coefficient $D_{lip} = 5\mu\text{m}^2/\text{sec}$. (36).

**Numerical simulations of the tubulation process.**

We tested this hypothesis with numerical simulations based on a continuum bilayer dynamical model, similar to that developed in (24) but capable of accessing the fully nonlinear regime (37). This model describes the bilayer by its midsurface shape and by a lipid density field for each monolayer. The viscoelastic response is determined by the stretching and curvature elasticity, and by the inter-monolayer friction and the membrane interfacial shear viscosity. The state of the bilayer is encoded in the model by the shape of the membrane surface $\Gamma$ and by lipid density fields in the outer (+) and inner (-) monolayer, $\rho^+$ and $\rho^-$ respectively. The elastic forces derive from the free energy

$$F = \frac{1}{2}\int_{\Gamma}\left\{ K_s\left[\left(\frac{\rho^+}{\rho_0}-1\right)^2 + \left(\frac{\rho^-}{\rho_0}-1\right)^2\right] + \kappa(2H)^2 \right\}dS \quad (3)$$

where $\Gamma$ denotes the total area of the membrane, $K_s$ is the monolayer stretching modulus as above, $\rho_0$ is the equilibrium lipid density, $\kappa$ is the bilayer bending modulus, $H$ is the membrane mean curvature, and any spontaneous curvature is neglected. The dissipative forces derive from the dissipation potential

$$D_{tot} = \int_{\Gamma}\left[\frac{b_m}{2}|\mathbf{v}^+ - \mathbf{v}^-|^2 + \mu(\mathbf{d}^+ : \mathbf{d}^+ + \mathbf{d}^- : \mathbf{d}^-)\right]dS \quad (4)$$

12

where $b_m$ is the inter-monolayer friction coefficient as above, $\mathbf{v}^+$ and $\mathbf{v}^-$ are the tangential velocity field vectors on each monolayer (37), $\mu$ is the monolayer interfacial viscosity, $\mathbf{d}^+$ and $\mathbf{d}^-$ are the rate of deformation tensors of each monolayer and "**:**" denotes the tensor contraction operator. To close the system, the balance of mass equations for each monolayer is used

$$\frac{d\rho^{\pm}}{dt} + \rho^{\pm} = \left(\mathrm{div}\mathbf{v}^{\pm} - 2v_n H\right) \quad (5)$$

where $v_n$ is the normal velocity. Since the protrusions are smaller than or comparable to the characteristic length $\kappa b_m / K_s \eta$ [38] (where $\eta$ is the viscosity of water), which is of the order of a few micrometers, the viscous forces due to the solvent are neglected. The applicability of such a simplifications are described in detail in our previous work (37, 39). The model is discretized under the assumption of axisymmetry, using B-Spline basis functions (37).

This model is able to capture the dynamics observed in our experiments since it allows for the variation of the lipid densities as a function of time and space as well as for the formation of arbitrary membrane shapes. As discussed in the previous section, the acid injection modifies the lipid packing by increasing the equilibrium density $\rho_0$ which places the system out of equilibrium. Thus the outer monolayer locally requires extra lipid to adopt the new equilibrium density, which promotes lipid flows on the both monolayers toward/from the modification area. The outer monolayer lipid flow is dragged by the membrane viscosity, due to the interfacial flow, and by intermonolayer friction due to the outer monolayer slippage on the inner monolayer. Simultaneously the new outer monolayer equilibrium density can be satisfied by the local membrane bending (invagination) which again derives an interfacial flow, partially slowed down by the membrane viscosity (Fig. 5 A, *left*) (37, 40). Similarly flip-flop can generate the same type of dynamics by depleting the outer monolayer and packing the inner monolayer (Fig. 5 A, *right*).

We chose as material parameters $\kappa = 10^{-19}$ J, $2K_s = 0.1$ J/m$^2$, $b_m = 10^9$ J s/m$^4$, and $2.\mu = 5.10^{-9}$ J s /m$^2$ [12, 41-44] The GUV was modeled as a deflated prolate vesicle of nominal radius 25 μm, similar to the vesicles in the experiments, and reduced volume $v = 0.85$ (Fig. 5 B).

A more detailed simulation of the process reported here would involve the advection/diffusion of H+ ions in the bulk, their reaction with the outer monolayer, and the subsequent effect on the bilayer dynamics. To investigate the plausibility of our hypothesis that asymmetric flip-flop is the major source for the different behavior of PG and CL GUVs, we focus on bilayer mechanisms, and build on our previous work to treat in a simplified way the diffusion of H$^+$ ions.

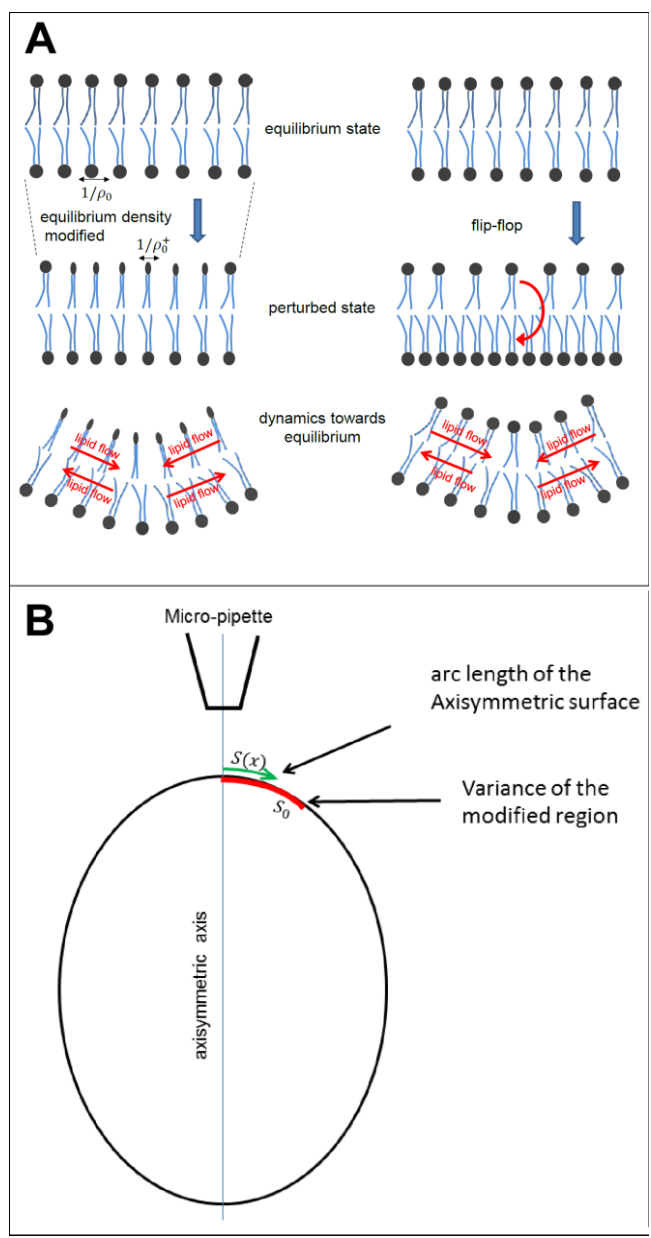

FIGURE 5 A. Schematic representation of the lipid density modifications, local shape changes and lipid flows occurring after acid injection on GUVs in the case of direct density change (left) and flip-flop (right). See text for details. B. Axially symmetric geometrical configuration used for simulations of acid-induced tubulation in CL or PG-containing GUVs. $S(x)$ is the arc length of the generating curve of the modified axisymmetric area. The red arc represents $S_0$, the standard deviation of the Gaussian function by which the lipid packing modification is applied.

We describe first the simulation protocol for the CL-containing membrane. The pH-induced lipid condensation is introduced without explicitly accounting for the local diffusion of $H^+$ ions due to acid injection and the subsequent protonation of lipid headgroups. A full treatment of this effect in the case of small deformations can be found in (12, 26). Instead, the equilibrium lipid density $\rho_0$ in the outer monolayer (+) in a small region of the vesicle $S_0$ was directly modified. To model such lipid condensation, the equilibrium density is abruptly increased by a constant amount for the duration of the injection time, i.e. in the interval (0;

14

$t_{inj}$). This simplification is justified because the acid-base equilibrium between the lipids and the injected acid is instantaneous, and the time taken by the pH profile to establish during the injection is of order 30 ms in our experiments, which is negligible compared to the injection time as shown in the Supporting Material, and in (26). Afterwards, this lipid density modification is gradually decreased in a time-scale $\tau$ commensurate to what we observe in experiments, in order to account for the diffusion of excess $H^+$ after the end of the injection. More specifically, the term ($\rho^+/\rho_0-1$) is replaced by($\rho^+/\rho_0^+ -1$) in the free energy (eq. 1), where $\rho_0^+ = \rho_0 [1 + \delta\rho_0 (x, t)]$, and

$$\delta\rho_0(x,t) = Ae^{-\max(0,t-t_{inj})/\tau} \cdot e^{-2S(x)^2/S_0^2} \quad (6)$$

where $x$ denotes a position in the generating curve of the axisymmetric surface, and $S(x)$ is the arc length of the patch subtended by $x$ and $S_0$, the standard deviation of the Gaussian function by which the lipid packing modification is applied (Fig. 5B). The Gaussian profile is motivated by the fact that the external concentration of $H^+$ is determined by a diffusion-dominated flow of the acidic solution from the micropipette (12-26). The duration of the injection phase was set to $t_{inj} = 10$ s, the relaxation time for the lipid density modification after injection to $\tau = 20$ s (given the complexity of theoretically estimating the rate of $H^+$ diffusion out of a semi-closed nanotube coated with $H^+$ binding groups, we choose a value commensurate with the CL tubule relaxation time), and the amplitude of the density modification to $A = 0.01$ (see previous paragraph). The variance of the Gaussian function used for the lipid packing modification $S_0$ was set to 7.5 μm, which is comparable to what was previously reported for the modified area due to the acidification (11).

To model the lipid flip from the outer to the inner monolayer in PG-containing membranes, a sink of lipids is added in the outer (+) monolayer, and a corresponding source of lipids in the inner (-) monolayer, proportional to the lipid density modification $\delta\rho_0$. This accounts for the fact that the protonated fraction of the lipids in (+) that has an enhanced probability of translocating to (-), is also responsible for the direct density increase and is therefore proportional to $\delta\rho_0$. This also accounts for the fact that when translocated to (-) these become unprotonated and do not come back to (+). For this purpose, the sink/source in the conservation of mass equation of the +/- monolayer is set to $B\delta\rho_0 (x, t)/A$. We chose $B = 2.10^{-4}$ s$^{-1}$ in agreement with (34).

Comparison of the simulations of direct lipid density modification alone (corresponding to CL) and direct lipid density modification plus flip-flop (corresponding to PG) is shown on Fig. 6 and movie S3. The introduction of flip-flop promotes a more rapid and intense tubulation with enhanced pearling. Interestingly, whereas tubulation due to direct density modification reaches a maximum at the time of end of injection, tubulation further increases afterwards when flip-flop is taken into account. This is due to the fact that additional flip-flop keeps occurring until the pH gradient is dissipated. Note however that this short delay between injection stop and maximum deformation may be difficult to detect experimentally since only the plane projection of the tube length is measured. As in the experiment, the tube formed in the presence of flip-flop relaxes more slowly than that formed with direct density modification alone (120 s vs. 30 s).

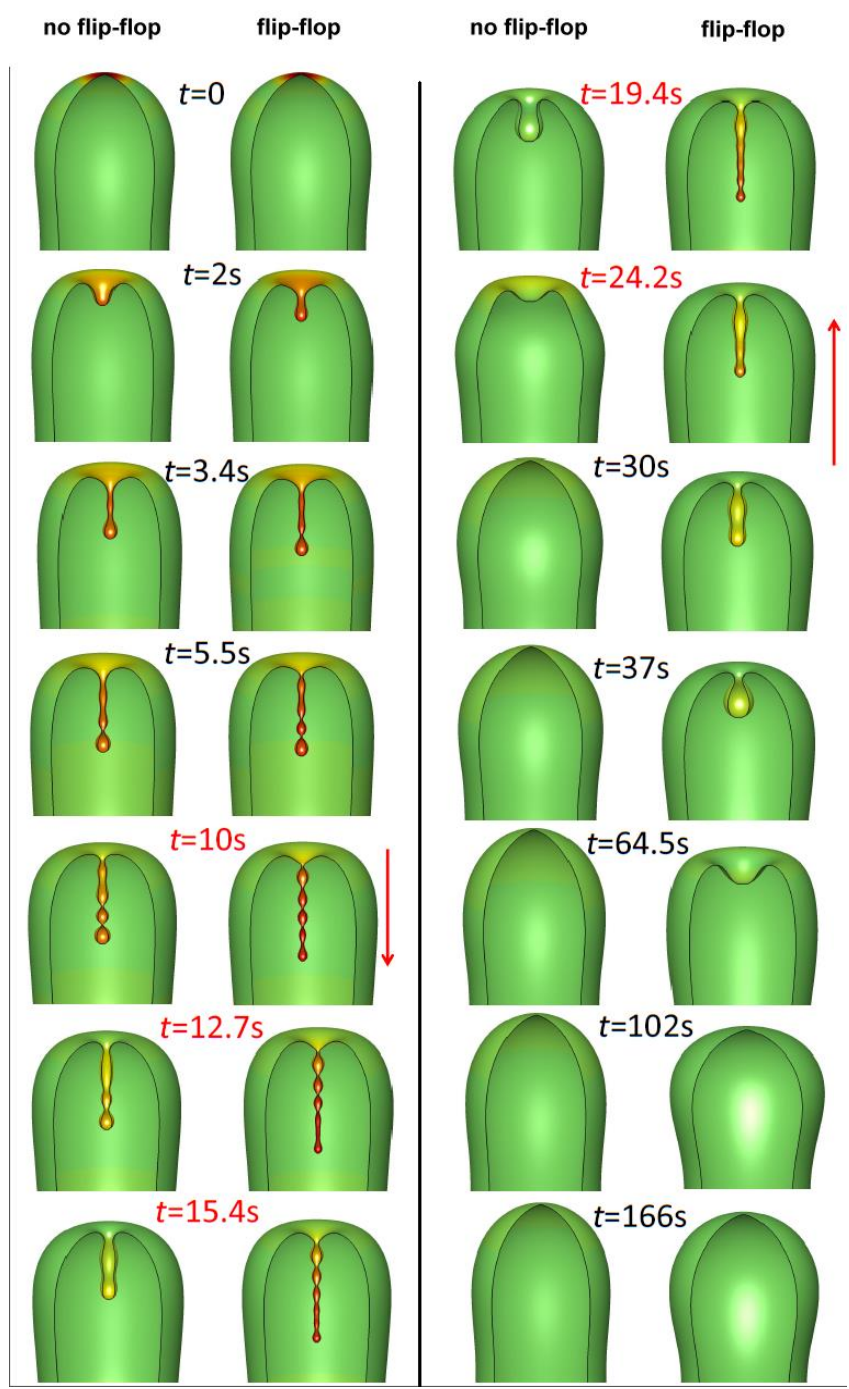

FIGURE 6  Results of simulation of tube formation after local acid injection on a GUV in the case of direct density change (corresponding to CL-containing GUVs, 1st and 3rd column) and in the case of direct density change and flip flop (corresponding to PG-containing GUVs, 2nd and 4rd column). The color maps represent the difference between the monolayer lipid densities $\rho^+$-$\rho^-$, where red corresponds to $A$ and green to 0. The *time values* and *arrows* in red correspond to the the time period where the acid diffuses and therefore the lipid density modification exponentially decreases. The corresponding movie is movie S3.

Fig. 7 shows the kinetics of growth and retraction of the tube corresponding to the simulation shown in Fig. 6. The tube length and kinetics both in the case of a direct equilibrium density change (corresponding to CL-containing GUVs) and in the case of a direct equilibrium density change plus flip flop (corresponding to PG-containing GUVs), although not identical to experiments, occur in similar time frames and with similar relative timescales (factor of 4 in relaxation times).

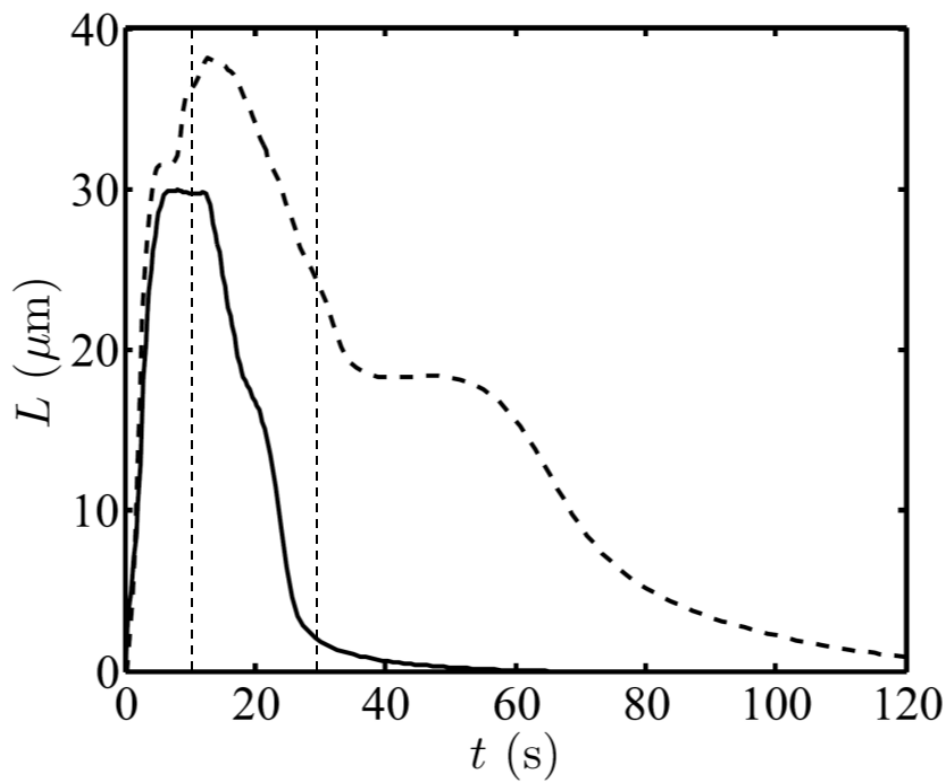

FIGURE 7 Dynamics of tube length growth derived from simulations and considering either only a direct equilibrium density change of the outer monolayer (*solid line*), corresponding to CL-containing GUVs, or a direct equilibrium density change and transbilayer sink/source flip flop (*dashed line*), (corresponding to CL-containing GUVs). The two vertical dotted lines correspond respectively to the end of the constant local density increase of the outer monolayer (i.e injection time) and the end of the relaxation of this density increase.

The evolution of dissipation power during the simulated tube growth and retraction was studied as a function of time (Fig. S3). Dissipation through intermonolayer friction due to monolayer sliding is the dominant process. However, transient increases in the fraction of total dissipation that arises from monolayer viscosity occur during the kinetics. As already pointed at in our previous simulations (37), these correspond to discontinuous events such as onset of tubulation, initial neck formation, appearance or retraction of pearls and final disappearance of neck and retraction. Such events are slowed down by the monolayer viscous drag. This is consistent with our recent theoretical work which describes in detail the coupling between out of plane membrane shape transformations and lateral monolayer viscous dissipation (40).

These simulations suggest that not only the more extended and pearled morphology of tubes formed with PG but also their slower retraction as compared to CL can be accounted for by a model in which local acidification promotes both direct change in lipid packing and inward PG flip-flop and in which tube retraction is limited by monolayer sliding and dissipation is dominated by intermonolayer friction. It is therefore plausible that, upon local exterior acidification of PC/PG-GUVs, the PG protonation-induced equilibrium density/spontaneous curvature mechanism and the PG flip-flop mechanism both contribute to the tubulation.

## CONCLUSION

The present study documents two possible 2 mechanisms involved in pH gradient-induced shape changes of charged lipid bilayers: (1) direct phospholipid packing (equilibrium density and spontaneous curvature) modifications; (2) asymmetric flip-flop (leading to bilayer density imbalance), both promoted by asymmetric protonation changes of lipid headgroups. Whereas the occurrence of the first mechanism only requires the presence of a protonable charged phospholipid, the second mechanism requires the occurrence of an uncharged form of this phospholipid, as well as a moderately sized headgroup. Apart from PG, phosphatidic acid (45) and phosphatidylinositol are likely candidates. Both mechanism lead to different relaxation properties, that influence the dynamics of shape changes.

We have shown that a pH gradient applied to a model membrane that mimics the mitochondrial inner membrane of CL synthase-deficient yeast mutant induces the formation of tubular structures. Our observation that the presence of PG in membranes does promote the formation of membrane tubules in the presence of a pH gradient provides an insight into the mechanism by which PG can partially substitute for CL in mitochondrial function. Indeed in the *cdr1*-Null mutant that lacks cardiolipin synthase and accumulates PG, tubules that may partially substitute to normal cristae may form in the IM, as found in PG-containing GUVs. This may explain why significant mitochondrial activity is present in this mutant despite the absence of CL (19-22). Furthermore, the occurrence of a flip-flop mechanism for PG may explain the particular temperature-sensitivity of mitochondrial function in the *cdr1-null* mutant. In ref. (33), it was found that PG flip-flop has an activation energy of 31kcal/mole, corresponding to a 3-fold increase of flip-flop rate every 7°C. This is likely to affect the mutant cristae morphology in vivo and therefore mitochondrial function. Therefore, this study gives further support to our suggestion of an important role of the lipid composition in the IM structure and function. Additionally, this study contributes to the general understanding of membrane tubulation, which occurs in processes as diverse as Golgi remodeling (46), endosome sorting (47), filopodia (48) and microvilli (49) formation, cell-cell contacts (50) and intercellular communications (51).

## SUPPORTING MATERIAL

Below: Three figures, details of local pH calculation, additional experimental results, simulation details, Supporting references.
On publisher's website: 3 videos.

## ACKNOWLEDGMENTS

**REFERENCES**

1. Farsad K. and P. De Camilli. Mechanisms of membrane deformation. 2003. *Curr. Opin. Cell. Biol.* 15:372-381.
2. Graham, T. R. and M. M. Kozlov 2010. Interplay of proteins and lipids in generating membrane curvature.*Curr. Opin. Cell Biol.* 22:430-436.
3. Bigay, J, and B. Antonny. Curvature, lipid packing, and electrostatics of membrane organelles: defining cellular territories in determining specificity. 2012. *Dev. Cell.* 23:886-895.
4. Miao L. , U. Seifert, M. Wortis, and H. G. Dobereiner. 1994. Budding Transitions of Fluid-Bilayer Vesicles: The Effect of Area-Difference Elasticity. *Phys. Rev.* E49, 5389-5407.
5. Seifert, U. 1997.Configurations of fluid membranes and vesicles. Adv. Phys. 46:13-137.
6. Mavcic, B., B Babnik, A Iglic, M Kanduser, T Slivnik, V Kralj-Iglic. 2004. Shape transformation of giant phospholipid vesicles at high concentrations of C12E8. *Bioelectrochemistry.* 63:183-187.
7. Staneva G., M. Seigneuret, K. Kouma, G. Trugnan, M. I. Angelova. 2005. Detergents induce raft-like domains budding and fission from giant unilamellar heterogeneous vesicles: a direct microscopy observation. *Chem. Phys. Lipids.* 136:55-66.
8. Itoh T. and  P. De Camilli. BAR, F-BAR (EFC) and ENTH/ANTH domains in the regulation of membrane-cytosol interfaces and membrane curvature. *Biochim. Biophys. Acta.* 2006. 1761:897-912.
9. Heuvingh, J., S. and Bonneau. 2009. Asymmetric oxidation of giant vesicles triggers curvature-associated shape transition and permeabilization. *Biophys J.* 97:2904-2912.
10. Tuller G, C Hrastnik, G Achleitner, U Schiefthaler, F Klein, G Daum. 1998. YDL142c encodes cardiolipin synthase (Cls1p) and is non-essential for aerobic growth of Saccharomyces cerevisiae. *FEBS Lett.*  421:15-18.
10. Yu Y. , J. A. Vroman, S.C.  Bae, S. Granick. 2010. Vesicle budding induced by a pore-forming peptide. *J Am Chem Soc.* 132:195-201.
11. Bitbol, A. F.; J. B Fournier; M. I. Angelova and N. Puff, 2011. Dynamical membrane curvature instability controlled by intermonolayer friction *J. Phys: Condens. Matter.* 23: 284102
12. Bitbol, A.-F., N.Puff, Y. Sakuma, M . Imai, J.-B. Fournier and M. I. Angelova, 2012. Lipid membrane deformation in response to a local pH modification: theory and experiments. *Soft Matter.* 8: 6073– 6082
13. Seigneuret, M. and  Devaux, P. 1984. ATP-dependent asymmetric distribution of spin-labeled phospholipids in the erythrocyte membrane: relation to shape changes. *Proc. Natl. Acad. Sci. USA*. 81, 3751-3755.
14. Farge, E. and Devaux, P. F. 1992. Shape changes of giant liposomes induced by an asymmetric transmembrane distribution of phospholipids. *Biophys J.* 61:347–357.

15. Farge, E. Increased Vesicle Endocytosis Due to an Increase in the Plasma Membrane Phosphatidylserine Concentration. *Biophys. J.* 69:2501-2506.

16. Logan, D.C. 2006. The mitochondrial compartment. *J. Exp. Bot.* 57: 1225–1243.

17. Mannella, C.A. 2006. Structure and dynamics of the mitochondrial inner membrane cristae. *Biochim. Biophys. Acta.* 1763: 542–548.

18. Janitor M., M. Obernauerová, S. D. Kohlwein and J. Subík. 1996. The pel1 mutant of Saccharomyces cerevisiae is deficient in cardiolipin and does not survive the disruption of the CHO1 gene encoding phosphatidylserine synthase. *FEMS Microbiol. Lett.* 140:43-47.

19. Jiang F., HS Rizavi, and M. L. Greenberg. 1997. Cardiolipin is not essential for the growth of Saccharomyces cerevisiae on fermentable or non-fermentable carbon sources. *Mol. Microbiol.* 26:481-491.

21. Chang, S. C, P. N. Heacock, C. J. Clancey and W. Dowhan. 1998. The PEL1 gene (renamed PGS1) encodes the phosphatidylglycero-phosphate synthase of Saccharomyces cerevisiae. *J. Biol. Chem.* 273:9829-9836.

22. Jiang, F., M. T. Ryan, M. Schlame, M. Zhao, Z. Gu, M. Klingenberg, N. Pfanner, and M. L. Greenberg. 2000. Absence of Cardiolipin in the crd1 Null Mutant Results in reased Mitochondrial Membrane Potential and Reduced Mitochondrial Function. *J. Biol. Chem.* 275:22387-22394.

23. Khalifat, N., N. Puff, S. Bonneau, J. B. Fournier, and M. I. Angelova. 2008. Membrane deformation under local pH gradient: mimicking mitochondrial cristae dynamics. *Biophys. J.* 95:4924-4933.

24. Khalifat, N.; Fournier, J. B.; Angelova, M. I.; Puff, N. 2011. Lipid packing variations induced by pH in cardiolipin-containing bilayers: the driving force for the cristae-like shape instability *Biochim. Biophys. Acta.* 1808, 2724– 2733

25. Fournier, J. B., N. Khalifat, N. Puff and M. I. Angelova. 2009. Chemically triggered ejection of membrane tubules controlled by intermonolayer friction. *Phys Rev Lett.* 102:018102.

26. Bitbol, A-F. and J-B. Fournier. 2013. Membrane properties revealed by spatiotemporal response to a local inhomogeneity. *Biochim. Biophys. Acta.* 1828:1241-1249.

27. Mathivet, L., S Cribier and P. F. Devaux. 1996. Shape change and physical properties of giant phospholipid vesicles prepared in the presence of an AC electric field. *Biophys J.* 70: 1112-1121.

28. Bagatolli L. A., T. Parasassi and E. Gratton 2000. Giant phospholipid vesicles: comparison among the whole lipid sample characteristics using different preparation methods: a two photon fluorescence microscopy study. *Chem Phys Lipids.* 105, 135-47.

29. Parasassi, T., G. De Stasio, G. Ravagnan, R. M. Rusch and E. Gratton. 1991. Quantitation of lipid phases in phospholipid vesicles by the generalized polarization of Laurdan fluorescence *Biophys. J.*, 60, 179– 189

30. Ninham B. W. and V. A. Parsegian. 1971. Electrostatic potential between surfaces bearing ionizable groups in ionic equilibrium with physiologic saline solution. *J Theor Biol.* 31:405-428.

31. Lewis R. N. and R.N. McElhaney. 2009. The physicochemical properties of cardiolipin bilayers and cardiolipin-containing lipid membranes. *Biochim. Biophys. Acta.* 1788:2069-2079.

32. Watts, A., K. Harlos, W. Maschke, and D. Marsh. 1978. Control of the structure and fluidity of phosphatidylglycerol bilayers by pH titration. *Biochim. Biophys. Acta* 510:63-74.

33. Hope, M. J., T. E. Redelmeier, K. F. Wong, W. Rodrigueza and P. R. Cullis. Phospholipid asymmetry in large unilamellar vesicles induced by transmembrane pH gradients. 1989. *Biochemistry.* 28: 1-4187.

34. Redelmeier, T. E., M. J. Hope and P. R. Cullis. 1990. On the mechanism of transbilayer transport of phosphatidylglycerol in response to transmembrane pH gradients. *Biochemistry*. 29:3046-3053.
35. Mui, B. L., Döbereiner, H. G., Madden T. D., Cullis, P. R. 1995. Influence of transbilayer area asymmetry on the morphology of large unilamellar vesicles. *Biophys J.* 69:930-941.
36. Almeida, P.F.F. and W.L.C.Vaz. 1995 Lateral Diffusion in Membranes. *In* Handbook of Biological Physics, Vol. 1. R. Lipowsky and E. Sackmann editors. Elsevier Science B.V. 306-357.
37. Rahimi, M. and Arroyo, M., 2012. Shape dynamics, lipid hydrodynamics and the complex viscoelasticity of bilayer membranes *Physical Review E*, 86: 011932.
38. Seifert, U. and S. A. Langer. 1993. Viscous Modes of Fluid Bilayer Membranes. *Europhys. Lett.* 23, 71-76.
39. Arroyo, M. and A. De Simone. 2009. Relaxation Dynamics of Fluid Membranes. *Phys. Rev. E* 79, 031915 (2009)
40. Rahimi,M., A. DeSimone and M. Arroyo. Curved fluid membranes behave laterally as effective viscoelastic media. 2013. *Soft Matter*. 9: 11033-11045.
41. Merkel, R., E.Sackmann and E.Evans, 1989. Molecular friction and epitactic coupling between monolayers in supported bilayers *J.Phys.France,* 50, 1535 -1555.
42. Rawicz, W., K.C.Olbrich,T. McIntosh, D.Needham and E.Evans. 2000 Effect of chain length and unsaturation on elasticity of lipid bilayers. *Biophys. J.* 79, 328-339.
43. T.Pott and P. Méléard. 2002. The dynamics of vesicle thermal fluctuations is controlled by intermonolayer friction *Europhys.Lett.* 59:87.
44. Horner, A., S. A. Akimov, and P. Pohl. 2013. Long and Short Lipid Molecules Experience the Same Interleaflet Drag in Lipid Bilayers. *Phys. Rev. Lett.* 110: 268101.
45. Eastman, S. J., M. J. Hope and P. R. Cullis. 1991. Transbilayer transport of phosphatidic acid in response to transmembrane pH gradients. *Biochemistry*. 30:1740-1745.
46. Ha K. D. , B. A. Clarke, and W. J. Brown. 2012. Regulation of the Golgi complex by phospholipid remodeling enzymes. *Biochim. Biophys. Acta.* 1821:1078-1088.
47. Breusegem, S. Y.and M. N. Seaman. 2014. Image-based and biochemical assays to investigate endosomal protein sorting. *Methods Enzymol.* 534:155-178.
48. Daniels, D. R. and M. S. Turner. Islands of conformational stability for filopodia. PLoS One. 208:e59010.
49. Maheshwari A. 2004. Role of cytokines in human intestinal villous development. *Clin Perinatol.*;31:143-155.
50. Okochi N., T. Okazaki, H. Hattori 2009. Encouraging effect of cadherin-mediated cell-cell junctions on transfer printing of micropatterned vascular endothelial cells. *Langmuir*. 25:6947-6953.
51. Gerdes, H. H. and R. N. Carvalho. 2008. Intercellular transfer mediated by tunneling nanotubes. *Curr Opin Cell Biol*. 20: 470-475.



\title{SUPPORTING MATERIAL \\Interplay of packing and flip-flop in local bilayer deformation. How phosphatidylglycerol could rescue mitochondrial function in a cardiolipin-deficient yeast mutant.}

\author{Nada Khalifat$^{\dag \ddag}$, Mohammad  Rahimi$^{\S}$, Anne-Florence  Bitbol$^{\S \parallel}$, Michel Seigneuret$^{**}$, Jean-Baptiste Fournier$^{**}$, Nicolas Puff$^{** \parallel\parallel}$, Marino Arroyo$^{\dag\dag}$, Miglena I. Angelova$^{** \parallel\parallel}$}

\address{$^{\dag}$UPMC Univ Paris 06, UMR 168, Institut Curie, 26 rue d’Ulm, 75248 Paris Cedex 05, France\\
$^{\ddag}$CNRS, UMR 168, Institut Curie, 26 rue d’Ulm, 75248 Paris Cedex 05, France\\
$^{\S}$Lewis-Sigler Institute for Integrative Genomics, Princeton University, Princeton, NJ 08544, USA\\
$^{\parallel}$Department of Physics, Princeton University, Princeton, NJ 08544, USA\\
$^{**}$Universit\'e Paris Diderot-Paris 7, Mati\`ere et Syst\`emes Complexes CNRS UMR 7057, Paris, France\\
$^{\parallel\parallel}$Universit\'e Pierre et Marie Curie, Paris, France\\
$^{\dag\dag}$Departament de Matem\`atica Aplicada III, LaC\`aN, Universitat Polit\`ecnica de Catalunya, 08034, Barcelona, Spain
}




\maketitle 

\Large \noindent \textbf{Determination of the pH at the membrane surface}\normalsize
\vspace{1cm}

\noindent
The pH on the GUV membrane during the microinjection of acid solution can be estimated as follows. First, the initial velocity $v_0$ of the acid solution when it just flows out of the micropipette into the fluid that surrounds the GUVs, i.e., the velocity right at the tip of the micropipette, is determined from the hydrodynamics inside the micropipette. This initial velocity $v_0$ is calculated as a function of the injection pressure, taking into account the geometry of the micropipette. Then, once the acid solution is out of the micropipette, we show that its dynamics is dominated by diffusion, which enables us to calculate the pH in the fluid that surrounds the GUVs, and hence the pH close to the GUV membrane. The diffusion calculation involves the quantity of acid injected per unit time from the micropipette, which acts as a source for the diffusion process. This quantity can be determined from $v_0$.

\section{Initial velocity $v_0$ of the acid solution}
\label{Sec1}

In order to estimate the velocity $v_0$ of the acid solution at the micropipette tip, we study the hydrodynamics in the micropipette.

Considering that the diameter of the micropipette varies slowly from the point where the injection pressure is imposed to the thin tip of the micropipette, we can approximate the shape of the micropipette as a succession of cylinders where the flow is a standard Poiseuille flow. The pressure drop $P$ in each of these cylinders is related to its diameter $\delta$ through $P\propto \delta^4/\ell$, where $\ell$ denotes the length of the cylinder~\cite{Guyon}. Hence, most of the pressure drop occurs in the thinnest region of the micropipette.

Measurements of the internal diameter of the micropipette tip, performed as described in~\cite{Schnorf94}, led to a value of $0.3\,\mu\mathrm{m}$  in our experiments. Moreover, direct observation under the microscope shows that the diameter of the micropipette is approximately constant along the first $\sim 10\,\mu\mathrm{m}$ closest to the tip, beyond which this diameter increases significantly. Hence, almost all the pressure drop occurs in this small cylinder region close to the tip of the micropipette. We thus approximate the velocity $v_0$ of the acid solution at the micropipette tip by the average velocity of a Poiseuille flow in a cylinder with diameter $\delta\approx0.3\,\mu\mathrm{m}$ and length $\ell\approx10\,\mu\mathrm{m}$ submitted to a pressure drop $P$ equal to the injection pressure, $P\approx20\,\mathrm{hPa}$. It reads
\begin{align}
v_0=\frac{P\delta^2}{32\eta \ell}\,,
\label{v0}
\end{align}
which yields $v_0\approx0.56\,\mathrm{mm/s}$ with the above-mentioned values.

In order to check the validity of this approximation, the velocity $v_0$ at the tip of the micropipette was also calculated assuming that the small cylinder considered here is followed by a cone and by another cylinder of larger diameter. The order of magnitude of the length and diameter of the cone and of the second cylinder were determined from the microscope observation of the micropipette, and the cone was treated as a succession of infinitesimal cylinders. This calculation yields a value of $v_0$ which is smaller than the above value by less than 4\%. Thus, the largest source of error in our estimate of $v_0$ does not arise from this approximation, but from our limited knowledge of the exact geometry of the tip of the micropipette, and in particular of the effective value of $\ell$, which could vary from one micropipette to another. In order to assess the uncertainty coming from this factor, we used both $\ell=10\,\mu\mathrm{m}$ and $\ell=5\,\mu\mathrm{m}$, and compared the values obtained. The final values of the pH were only marginally affected (see below).

\section{Diffusion of the acid solution}
\label{secdiff}

Once the acid solution is out of the pipette, its dynamics is determined by diffusion and by Stokes hydrodynamics. The relative importance of these two effects is characterized by the P\'eclet number $\mathrm{Pe}=vz_0/D$, where $v$ is the flow velocity, while $D$ is the diffusion coefficient of hydrochloric acid in water, $D=2/(1/D_{\mathrm{H}^+}+1/D_{\mathrm{Cl}^-})=3347\,\mu\mathrm{m^2/s}$ at infinite dilution and at $20 \,^{\circ}\mathrm{C}$~\cite{Cussler}, and $z_0$ is a characteristic distance of the flow, which is taken as the distance between the membrane and the micropipette tip, i.e., about $10\,\mu\mathrm{m}$. Although the actual characteristic length scale over which the velocity varies is in fact smaller (see below), we take this value in order not to risk underestimating the P\'eclet number. At the very tip of the micropipette, where $v=v_0$, $\mathrm{Pe}\approx 1.7$, which implies that both hydrodynamic flow and diffusion are initially important. However, away from the micropipette, the velocity of the flow decreases, and hence diffusion becomes dominant. To demonstrate this, we first assume the opposite, i.e. that hydrodynamics is dominant over diffusion, thereby overestimating the flow velocity. In this case, the flow corresponds to a Stokes jet~\cite{Birkhoff, Lakshmi66}. Flow conservation (or, equivalently, more complete Stokes jet dynamics, see~\cite{Birkhoff, Lakshmi66}) entails that at a distance $L$ from the source, the flow velocity is $v\approx v_0 \delta^2/(16L^2)$. For $L=1\,\mu\mathrm{m}$, we obtain $v\approx3.2\,\mu\mathrm{m}/s$, which gives a P\'eclet number $\mathrm{Pe}=vz_0/D\approx 10^{-2}$, in contradiction with the initial assumption. This shows that once the acid solution is more than $1\,\mu\mathrm{m}$ away from the micropipette tip, its dynamics is dominated by diffusion. This in line with our previous experimental work (Fig.~6 of Ref.~\cite{Bitbol12_guv}) in which the injection of a fluorescent substance under identical conditions led to the observation of a quasi-spherical fluorescent ``cloud'' characteristic of diffusion.

Thus, in order to calculate the pH profile in the fluid that surrounds the GUVs, we can consider that the acid solution simply diffuses from the micropipette tip. Given the cylindrical symmetry about the axis of the micropipette, we use polar coordinates, with origin the projection of the micropipette tip on the membrane. The coordinates of the micropipette tip are then $(r=0,z_0)$. The concentration $C$ of protons can be obtained by solving the (three-dimensional) diffusion equation 
\begin{equation}
\partial_t C-D\nabla^2 C=\Sigma\,.
\label{DE}
\end{equation}
where $\Sigma$ is the source term arising from microinjection. For a constant injection flow from the point $(r=0,z_0)$ starting at time $t=0$, and stopping at time $T$, it reads at position $(\bm{r},z)$ and time $t$
\begin{equation}
\Sigma(\bm{r},z,t)=\Sigma_0 \,\delta(\bm{r})\,\delta(z-z_0)\,\bm{1}_{[0,T]}(t)\,,
\label{S}
\end{equation}
where $\bm{1}_{[0,T]}$ is the indicator function of the interval $[0,T]$, while $\Sigma_0$ is a constant, and $\bm{r}$ denotes the in-plane radial vector of length $r$. Here, we are interested in calculating the pH on the membrane during the injection phase, i.e., for $t<T$. However, solving the diffusion problem also gives the acid concentration profile after the injection, i.e., during the relaxation of the membrane deformation (see Ref.~\cite{Bitbol12_guv}).

\paragraph{Determination of $\Sigma_0$.}
In a volume $V$ which contains the micropipette tip, the amount of acid which appears in the fluid due to injection during a short time interval $dt$ reads
\begin{equation}
dt\int_V  d\bm{r}dz\,\Sigma(\bm{r},z,t)=\Sigma_0 \,dt=\frac{\pi}{4} C_\mathrm{int}\delta^2 v_0\, dt\,.
\end{equation}
where $C_\mathrm{int}$ is the acid concentration inside the micropipette, i.e. 100 mM in our experiments. Consequently, we obtain
\begin{equation}
\Sigma_0=\frac{\pi}{4} C_\mathrm{int}\delta^2 v_0\,. \label{Szero}
\end{equation}

\newpage

\paragraph{Two simple geometries.}

\begin{figure}[h]
\centering
\subfloat[$z_0\gg R$]{\includegraphics[width=0.08\textwidth]{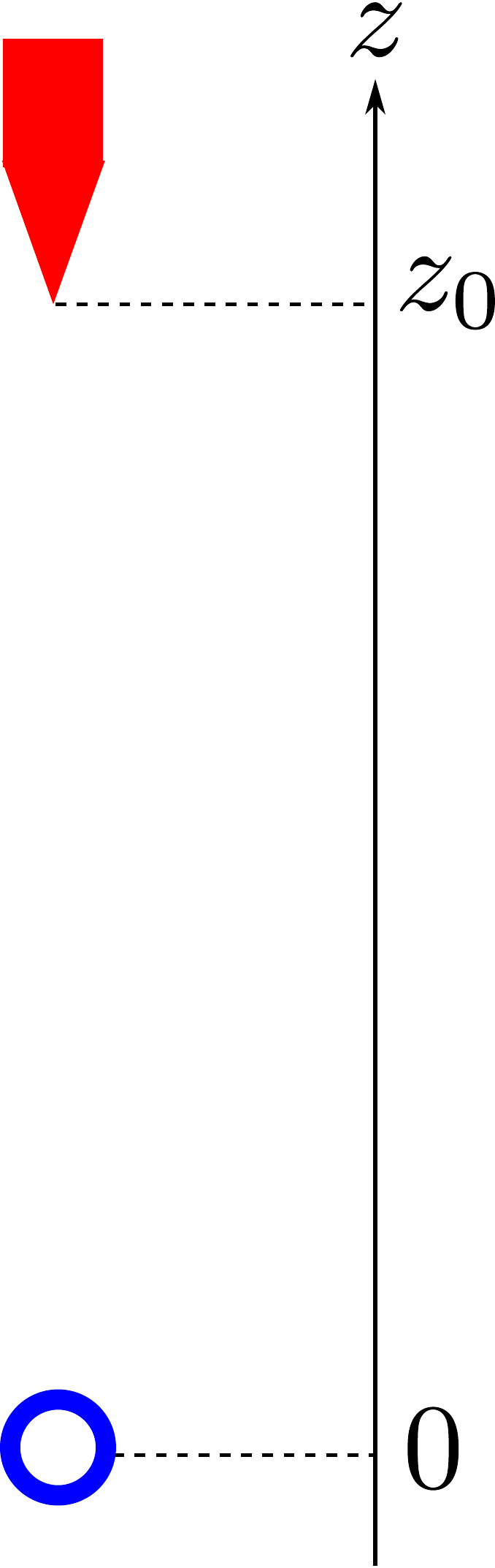}}
\quad\quad\quad\quad
\quad\quad\quad\quad
\subfloat[$z_0\ll R$]{\includegraphics[width=0.3\textwidth]{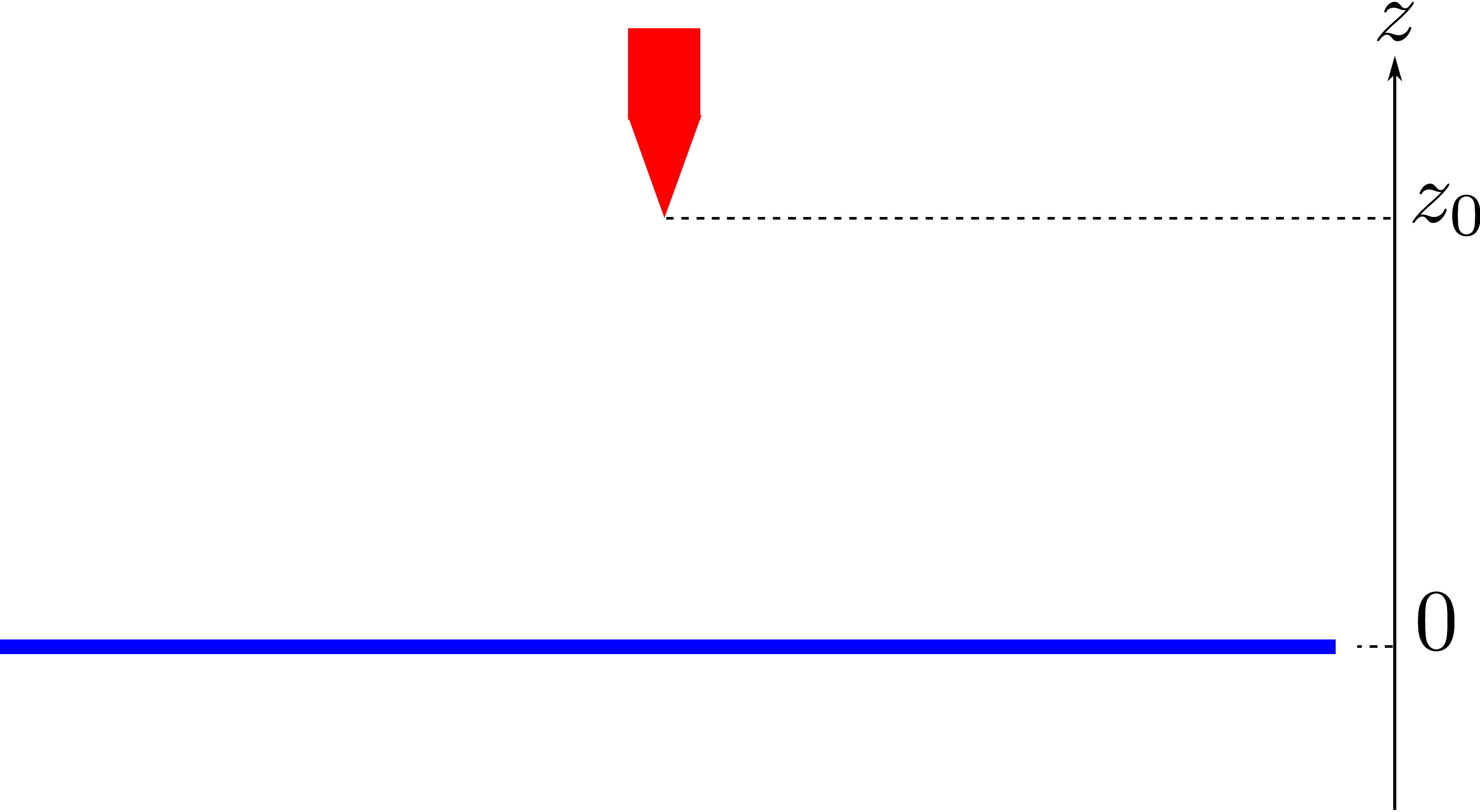}}
\caption[]{Two simple geometries. Red: micropipette, blue: membrane of the GUV. Denoting by $R$ the radius of the GUV, and by $z_0$ the height of the micropipette tip measured from the membrane, these geometries correspond to the asymptotic limits $z_0\gg R$ and $z_0\ll R$.\label{dist}}
\end{figure}

It is possible to obtain analytical expressions of the proton concentration profile $C(r,z,t)$ in two simple geometries denoted by (a) and (b), and defined in Fig.~\ref{dist}.

In geometry (a), i.e., for $z_0\gg R$, we can consider that the acid diffuses freely in an infinite volume in order to calculate the proton concentration $C$. Conversely, in geometry (b), i.e., for $z_0\ll R$, we can consider that the membrane imposes a Neumann boundary condition to $C$: 
\begin{equation}
\partial_z C\left(r,z=0,t\right)=0\,.
\label{nbc}
\end{equation}
This boundary condition, corresponding to complete reflection on the membrane, is fully valid in the stationary state, where the membrane is at chemical equilibrium with the surrounding solution, so that there are as many protons per time unit that adsorb onto the membrane (by reacting with a lipid) as protons that desorb from the membrane. In the transient state at the beginning of the injection, taking this boundary condition amounts to neglecting the excess protons that react with the membrane, which is legitimate given that the membrane is a surface and that the fraction of lipids that react with protons is assumed to remain small. As shown below, the acid concentration on the vesicle can be considered constant during most of the injection phase.

\paragraph{Resolution of the diffusion problem.}

The solution to the diffusion equation Eq.~(\ref{DE}) can be written as:
\begin{align}
C\left(r,z,t\right)&=\int_0^{t} \!\!dt' \int_{\mathbb{R}^2} \!\!\!\!d\bm{r'} \int_0^{+\infty}\!\!\!\!\!\!\!\!dz'\,\Sigma\left(\bm{r'},z',t'\right)\,G\left(|\bm{r}-\bm{r'}|,z,z',t-t'\right)\nonumber\\
&=\Sigma_0\int_0^{\mathrm{min}(t,\,T)} dt'\,G\left(r,z,z_0,t-t'\right)\,,
\label{cint}
\end{align}
where $G$ is the causal Green function of the diffusion equation which verifies the appropriate boundary conditions. \\
In geometry (a), it is simply the infinite-volume causal Green function of the diffusion equation:
\begin{equation}
G_\infty \left(r,z,z_0,t\right)=\frac{\theta(t)}{\left(4\,\pi\,D\,t\right)^{3/2}}\,\exp\left(-\frac{r^2+(z-z_0)^2}{4\,D\,t}\right)\,,
\label{G2}
\end{equation}
where $\theta$ is the Heaviside function. \\
In geometry (b), the boundary condition Eq.~(\ref{nbc}) can be accounted for using the method of images~\cite{Alastuey}, yielding:
\begin{equation}
G\left(r,z,z',t\right)=G_\infty \left(r,z,z',t\right)+G_\infty \left(r,z,-z',t\right)\,.
\label{G1}
\end{equation}
Using Eqs.~(\ref{cint}), (\ref{G2}) and (\ref{G1}), it is possible to obtain explicit analytical expressions for $C(r,z,t)$ in both geometries~\cite{Bitbol12_guv, Bitbol13_theodiff}. 

\paragraph{Steady-state concentration on the GUV membrane.}
Here, we are interested in calculating the pH imposed on the GUV membrane during the injection, which lasts $T\approx30\,\mathrm{s}$ in our experiments. Since the concentration of protons is determined by diffusion of the acid solution from the micropipette tip, this concentration can be considered to have reached steady state at the point of coordinates $(r,z)$ if the time $t$ from the beginning of the injection verifies $t\gg (r^2+(z-z_0)^2)/D$. For instance, on the membrane in front of the micropipette, it corresponds to $t\gg z_0^2/D\approx 30\,\mathrm{ms}$, where we have used the above-mentioned values for $z_0$ and $D$. Since this threshold is much smaller than the total injection time $T$, the relevant value of the proton concentration during the injection is that at steady-state. \\
In the (a) geometry, the steady-state concentration at the membrane surface (i.e., at $z=0$) reads
\begin{equation}
C^\mathrm{(a)}\left(r\right)=\frac{\Sigma_0}{4\,\pi\,D}\,\,\frac{1}{\sqrt{r^2+z_0^2}}\,,
\label{cr2mm}
\end{equation}
and in the (b) geometry, it reads
\begin{equation}
C^\mathrm{(b)}\left(r\right)= 2\,C^\mathrm{(a)}\left(r\right)\,.
\label{cr2ee}
\end{equation}
Note that the twofold difference between cases (a) and (b) directly arises from the method of images~\cite{Alastuey}.

\section{pH in front of the micropipette during the experiments}

Using Eqs.~(\ref{cr2mm}) and (\ref{cr2ee}) together with Eqs.~(\ref{Szero}) and (\ref{v0}), the acid concentration on the membrane during the experiments can be estimated. We calculate it for the steady state of the injection phase (as explained above), and in front of the micropipette (in $r=0$), because it is where the pH is lowest. 

In our experiments, the distance between the micropipette tip and the membrane is $z_0=10\,\mu\mathrm{m}$, and the radius $R$ of the GUV is about $25\,\mu\mathrm{m}$. Hence, the actual pH is expected to be comprised between those calculated for the limiting geometries (a) and (b). We carried out the calculations in both cases.

Besides, we took into account the uncertainty on the injection pressure (15-25 hPa), as well as on the effective value of $\ell$, (i.e. the 5-10 $\mu\mathrm{m}$ length over which the micropipette has its smallest diameter $\delta=0.3\,\mu\mathrm{m}$). 
Calculating the pH from Eqs.~(\ref{cr2mm}) and (\ref{cr2ee}) for all these sets of values, yielded two pH extremes, namely 4.4 and 5.2.
Hence, it can be concluded that during the microinjection, the pH at the GUV membrane surface, in front of the micropipette, is
\begin{equation}
\mathrm{pH}=4.8\pm0.4\,.
\end{equation}

Since this range of values is more that 2 units below the pKa of Hepes, the effect of the buffer can be neglected.

\newpage
\Large \noindent \textbf{Additional experimental results}\normalsize
\vspace{1cm}

\begin{figure}[h]
\centering
\includegraphics[width=0.9\textwidth]{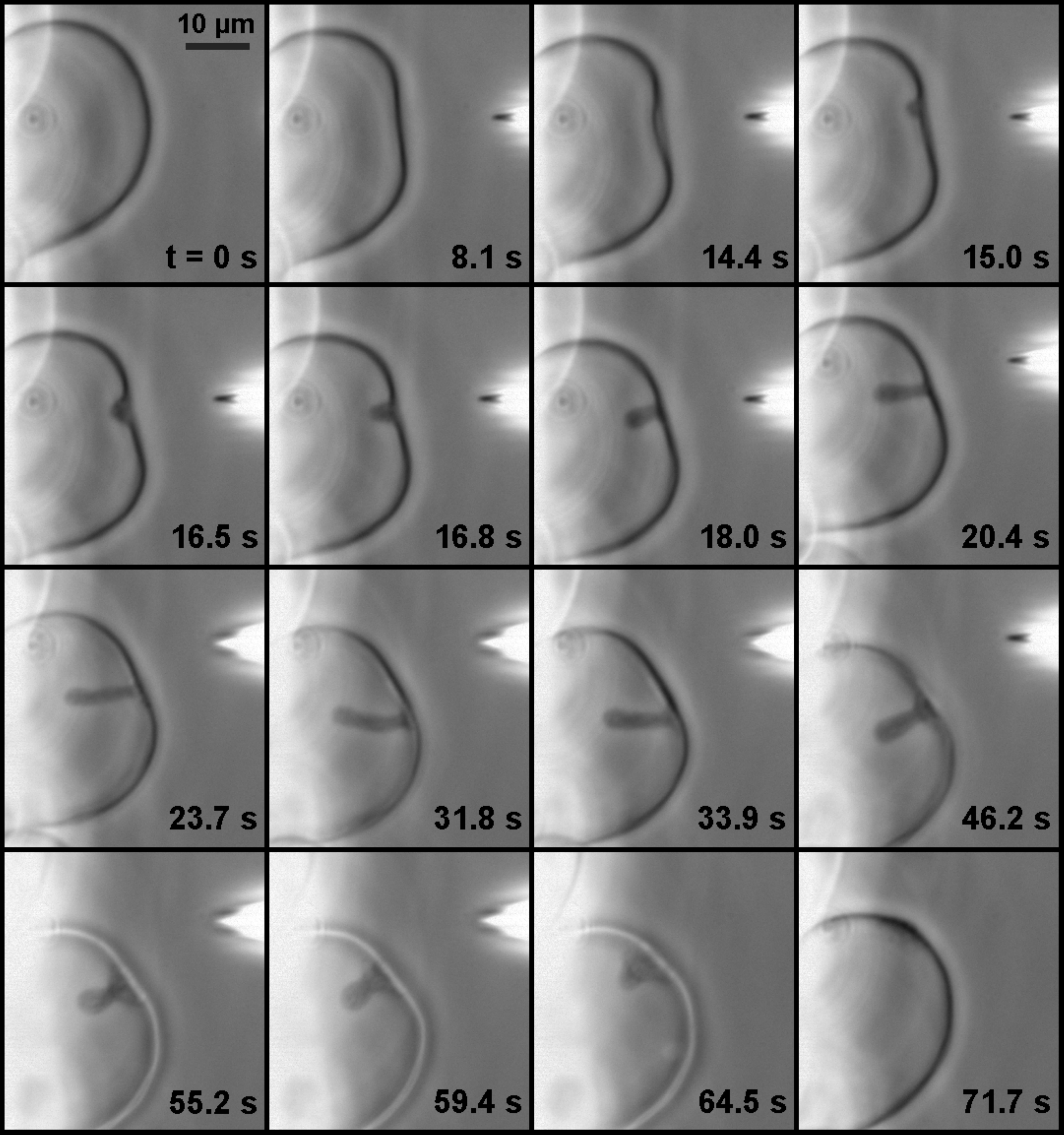}
\caption[]{Local membrane invagination triggered on a GUV composed of PC/CL 90:10 mol/mol by a local pH decrease performed using a micropipette delivering an acidic solution (100 mM HCl, pH 1.6) in buffer A at pH 8. The acid microinjection is started at t=0 and ended at t=31.8 s. The corresponding movie is movie S1.}
\end{figure}

\newpage
\Large \noindent \textbf{Details of simulation results}\normalsize
\vspace{1cm}

\begin{figure}[h]
\centering
\includegraphics[width=0.45\textwidth]{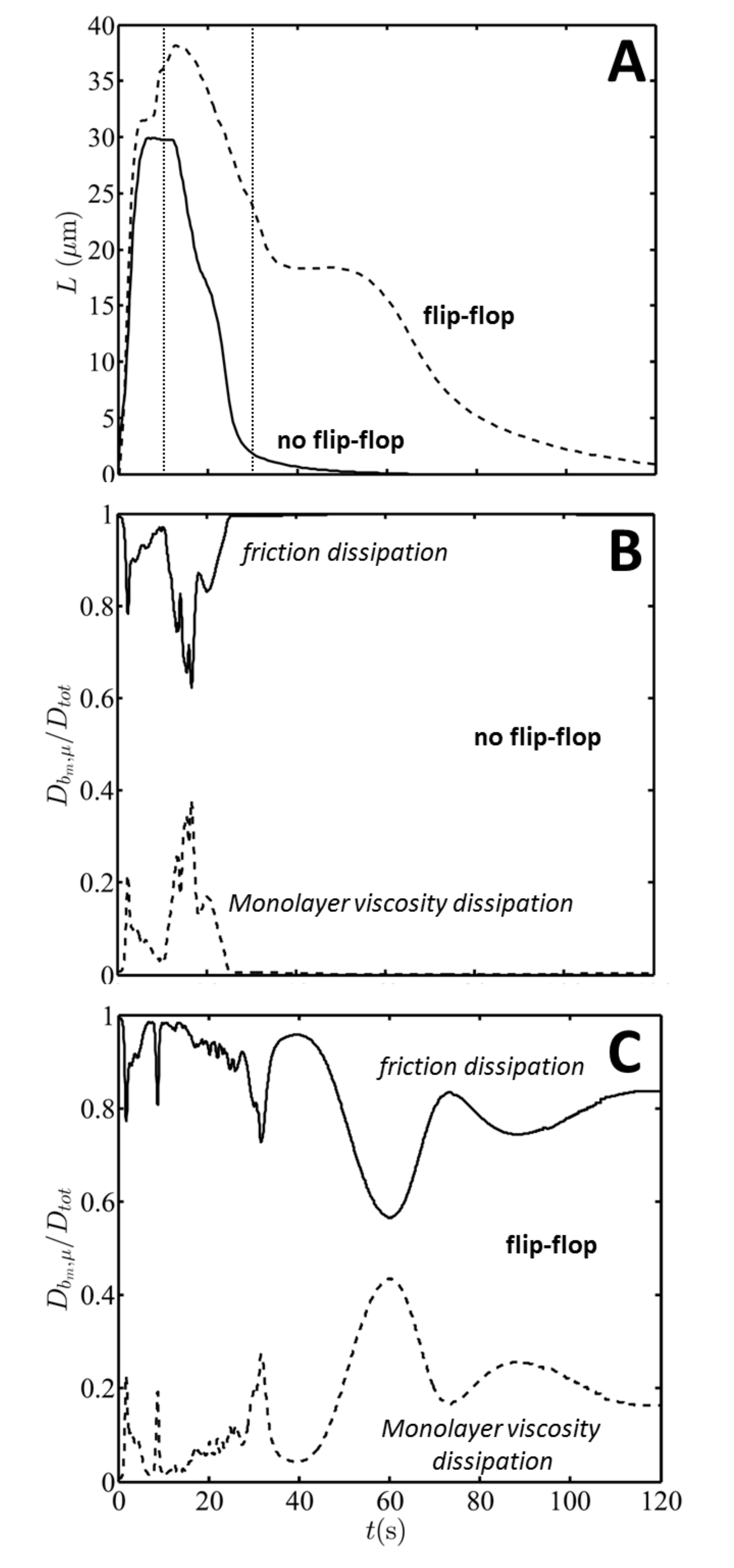}
\caption[]{\textbf{A.} Dynamics of the growth of tube length $L$ derived from simulations and considering either only a direct equilibrium density change of the outer monolayer (solid line), corresponding to CL-containing GUVs, or both a direct equilibrium density change and transbilayer sink/source flip flop (dashed line), corresponding to CL-containing GUVs. The two vertical dotted lines correspond respectively to the end of the constant local density increase of the outer monolayer (i.e. injection time) and the end of the relaxation of this density increase.  \protect\\ \textbf{B, C.} Dynamics of the fractions of the total dissipated power $D_{tot}$ arising from intermonolayer friction ($D_{b_m}$, solid line) and monolayer viscosity dissipation ($D_{\mu}$, dotted line) in the case of only a direct equilibrium density change of the outer monolayer (\textbf{B}), or of both a direct equilibrium density change and transbilayer sink/source flip flop (\textbf{C}).}
\end{figure}






\begin{thebibliography}{8}
\providecommand{\url}[1]{\texttt{#1}}
\providecommand{\urlprefix}{ }

\bibitem[Guyon et~al.(2001)Guyon, Hulin, Petit, and Mitescu]{Guyon}
Guyon, E., J.-P. Hulin, L.~Petit, and C.~D. Mitescu, 2001.
\newblock Physical Hydrodynamics.
\newblock Oxford University Press.

\bibitem[Schnorf et~al.(1994)Schnorf, Potrykus, and Neuhaus]{Schnorf94}
Schnorf, M., I.~Potrykus, and G.~Neuhaus, 1994.
\newblock {Microinjection technique: routine system for characterization of
  microcapillaries by bubble pressure measurement}.
\newblock \textit{Exp. Cell Res.} 210:260--267.

\bibitem[Cussler(2009)]{Cussler}
Cussler, E.~L., 2009.
\newblock Diffusion -- Mass transfer in fluid systems.
\newblock Cambridge University Press.

\bibitem[Birkhoff and Zarantonello(1957)]{Birkhoff}
Birkhoff, G., and E.~H. Zarantonello, 1957.
\newblock Jets, wakes and cavities.
\newblock Academic Press.

\bibitem[Lakshmi({1966})]{Lakshmi66}
Lakshmi, S.~V., {1966}.
\newblock {Asymmetric creeping flow from an orifice in a plane wall}.
\newblock \textit{Appl. Sci. Rev} {17}:{355--358}.

\bibitem[Bitbol et~al.(2012)Bitbol, Puff, Sakuma, Imai, Fournier, and
  Angelova]{Bitbol12_guv}
Bitbol, A.-F., N.~Puff, Y.~Sakuma, M.~Imai, J.-B. Fournier, and M.~I. Angelova,
  2012.
\newblock {Lipid membrane deformation in response to a local pH modification:
  theory and experiments}.
\newblock \textit{Soft Matter} 8:6073--6082.

\bibitem[Alastuey et~al.(2008)Alastuey, Magro, and Pujol]{Alastuey}
Alastuey, A., M.~Magro, and P.~Pujol, 2008.
\newblock Physique et outils math\'ematiques -- m\'ethodes et exemples.
\newblock EDP Sciences and CNRS Editions.

\bibitem[Bitbol and Fournier(2013)]{Bitbol13_theodiff}
Bitbol, A.-F., and J.-B. Fournier, 2013.
\newblock {Membrane properties revealed by spatiotemporal response to a local
  inhomogeneity}.
\newblock \textit{Biochim. Biophys. Acta -- Biomembr.} 1828:1241--1249.

\end{thebibliography}

 \bibliographystyle{biophysj}








\end{document}